\documentclass[draft,showpacs,preprintnumbers,amsmath,amssymb]{revtex4-1}

\usepackage{graphicx}
\usepackage{color}

\begin{document}
\title{Linearized modified gravity theories with a cosmological term: advance of perihelion and deflection of light}

\begin{abstract}
Two different ways of generalizing Einstein's general theory of relativity with a cosmological constant to Brans-Dicke type scalar-tensor theories are investigated in the linearized field approximation.  In the first case a cosmological constant term is coupled to a scalar field linearly whereas in the second case an arbitrary potential plays the role of a variable cosmological term. We see that the former configuration leads to a massless scalar field whereas the latter leads to a massive scalar field. General solutions of these linearized field equations for both cases are obtained  corresponding to a static point mass. Geodesics of these solutions are also presented and solar system effects such as the advance of the perihelion, deflection of light rays and gravitational redshift were discussed. In general relativity cosmological constant has no role  on these phenomena. We see that for the Brans-Dicke theory the cosmological constant has also no  effect on these phenomena. This is because solar system observations require very large values of the Brans-Dicke parameter and the correction terms to these phenomena becomes identical to GR for these large values of this parameter. This result is also observed for the theory with arbitrary potential if the mass of the scalar field is very light. For a very heavy scalar field, however, there is no such limit on the value of this parameter and there are ranges of this parameter where these contributions may become relevant in these scales.  Galactic and intergalactic dynamics is also discussed for these theories at the latter part of the paper with similar conclusions.
 
\end{abstract}

\pacs{04.25.Nx,04.50.Kd}

\author{Hatice \"Ozer}
\email{hatice.ozer@istanbul.edu.tr}
\affiliation{Department of Physics, Faculty of  Sciences, Istanbul University,  Istanbul, Turkey}
\author{\"Ozg\"ur Delice}
\email{ozgur.delice@marmara.edu.tr}
\affiliation{Department of Physics, Faculty of Arts and Sciences, Marmara University,  Istanbul, Turkey}
\date{\today}

\maketitle

\section{Introduction}

The remarkable observation at the end of 20th century showed that we live in an accelerating universe  \cite{darken1,darken2,darken3}. According to  well tested theory of gravitation, namely Einstein's  general relativity (GR)  theory, this cosmic accelerating expansion is caused by a mysterious component  of the universe called dark energy. The most clear  candidate  of dark energy  is Einstein's cosmological constant, since, this observed behavior of the universe is  compatible with a very small positive cosmological constant, i.e., $\Lambda\sim 10^{-52}\,m^{-2}$ \cite{darken1,darken2,darken3}. Another candidate is quintessence in the form of a minimally coupled scalar field which varies slowly along its potential \cite{quint1,quint2,quint3}. For a review on quintessence, see, for example, see \cite{quint4}.  Considering thus the fact that we live in an asymptotically de Sitter universe,  it might be reasonable to investigate the  possible effects of  a positive cosmological constant into local and global behavior of the universe. Therefore, it is logical to investigate whether such a cosmological constant, despite its smallness, affects the local gravitational phenomena such as bending of light from distant objects or the advance of perihelion of objects in bound orbits. 

 Alternative theories of GR has been a very popular field of research, especially over recent decades. There are several theoretical or observational motivations that exist for this active field of research. One of them is  to understand the mathematical structure, physical predictions and  behaviour of GR by studying its alternatives. Another one is the quantization of gravitational interaction, and the fact that it may  require some modifications to  GR \cite{utiyama,Stelle1}. One more reason is the idea of unification of fundamental interactions, generalizing the Kaluza-Klein idea of unifying gravity and electromagnetism \cite{Kaluza,Klein} into all interactions, such as string theory \cite{String}. Such attempts require the ideas of the existence of extra dimensions  and compactification. Such a   compactification of higher dimensional theories into four dimensions usually  produces a scalar field called dilaton into the four dimensional effective theories \cite{Kaluza,Klein,String}. Apart from these, modified gravity theories,  such as $f(R)$ theory, are also popular to investigate the possibility that the accelerating universe may be explained by large scale modifications to GR, without needing a dark energy.  We refer to the latest reviews for the further motivations and developments of these theories \cite{modgr1,modgr2,modgr3,modgr4,modgr5}.  

 Brans-Dicke (BD) theory  \cite{ST1,ST2,BD} is one of the most simple modifications to GR and usually considered as a  suitable test bed for investigating the effects of possible modifications to GR. After its presentation more than half a decade ago, the properties and outcomes of this theory is investigated in great detail \cite{Brans,Fujii-maeda,Faraonibook}. For example, its  weak field   solution  for  point particle is obtained  and two most interesting weak field phenomena, namely the perihelion precession of Mercury and light deflection by the Sun is investigated in the original papers of this theory \cite{BD,Brans}. In its original form, as we will discuss in the next section, BD theory does not involve neither a cosmological constant or  a potential term. However, in the later years those extensions were also discussed, mostly in the cosmological scheme.

In this paper, we investigate weak field solutions of theories which generalizes Einstein's general relativity with a cosmological constant to the Brans-Dicke type scalar-tensor theory. Since, as we will discuss in the next section, this generalization can be made, at least, in two different ways, we will consider both cases, separately. There are many works  considering  weak field solutions, properties of these solutions and astrophysical implications for different modified gravity theories \cite{Pechlaner,Wagoner,Stelle,Will,Steinhardt-will,Barros,olmo,olmo11,olmo1,Perivolaropoulos,Berry-gair,Alsing-berti-will,Hohmann}. However, in most of the works, except for example \cite{olmo,olmo11,olmo1}, asymptotic flatness is assumed. In \cite{olmo11} a post-Newtonian extension of BD theory with a potential was presented. Our motivation in this paper is to investigate the weak field solutions of BD theory in the presence of an asymptotically de Sitter background. This will enable us to  shed light into the effects of  background curvature on the dynamics of the space time in the presence of positive cosmological constant in these theories. We transform the linearized field equations in a known suitable gauge which  makes scalar and tensor equations decouple from each other and  makes it  easier to obtain the solutions.  We will solve these equations for a static point particle in the coordinates where this gauge is valid  for both cases and transform the obtained solutions into isotropic or Schwarzschild-like coordinates where this gauge is not valid. In order to obtain physical properties of  these solutions, we will discuss the geodesics of both solutions in Schwarzschild type coordinates. Advance of the  perihelion of test particles around this point particle, deflection of light rays by this point particle in the presence of a curvature background and also the gravitational redshift and galaxy rotation curves and intergalactic dynamics will be discussed for both solutions. Contribution of the mass of the source, the cosmological term or the minimum of the potential to these phenomena will be derived using appropriate methods. The paper is organized as follows. In the next section, we will discuss two different ways of generalizing  GR with a cosmological constant  to BD theory and obtain the  weak field equations in a chosen gauge for both cases. In section (III) we will present a static point particle as a source, solve the field equations for both cases in the chosen gauge and transform the solutions  to the isotropic and Schwarzschild type coordinates. In section (IV) we will obtain radial geodesic equations in Schwarzschild coordinates.   We will investigate  solar system effects such as the advance of perihelion in  section (V), deflection of light rays in section (VI) and gravitational redshift in section (VII) for both of the theories. Galactic and intergalactic dynamics is considered in section (VIII). The paper ends with a brief discussion.

\section{Weak Field Equations}
According to Einstein's general theory of relativity (GR), the gravitational phenomena can be explained by the following action
\begin{equation}\label{EH-action}
\mathcal{S}_{GR \Lambda }=\int \sqrt{|g|} d^4x\left[  \frac{1}{2\kappa} \left( R-2\Lambda \right) +\mathcal{L}_{matter} \right].
\end{equation}
Here, Einstein's famous modification of adding a cosmological constant  term to the action is already included. We may call this theory as GR$\Lambda$ theory.  Here $\kappa=8\pi G/c^4$ is the gravitational coupling constant, $T_{\mu\nu}$ is the energy-momentum tensor,   $R$ is the Ricci scalar and $\Lambda $ is the cosmological constant term and we choose the units where $c=G=1$ in this paper. One of the most studied alternative of GR is the Brans-Dicke (BD) scalar-tensor theory \cite{BD}, where the Newton gravitational constant $\kappa$ is replaced by a scalar function as $\kappa\rightarrow 8 \pi\phi^{-1}$ together with addition of a kinetic term for this scalar field coupled by a dimensionless constant known as the BD parameter $\omega$. In the original derivation of the BD theory, cosmological constant  $\Lambda$ is set to zero. However, if one wants to extend GR theory with a cosmological constant to scalar-tensor theories, the most straightforward way is to replace $\kappa\rightarrow 8 \pi\phi^{-1}$ in action (\ref{EH-action}), similar to the original BD theory. This yields  the following action in Jordan frame
\begin{equation}\label{BD-action}
\mathcal{S}_{BD\Lambda}=\int \sqrt{|g|} d^4x\left\{  \frac{1}{16 \pi} \left[ \phi \left(  R-2\Lambda \right)-\frac{\omega}{\phi}g^{\mu\nu}\partial_\mu \phi \,\partial_\nu \phi  \right] +\mathcal{L}_{matter} \right\}.
\end{equation}
This action where all curvature related terms $R$ and $\Lambda$ is coupled with scalar field $\phi$ in the same manner, is known as the Brans-Dicke theory with a cosmological constant  \cite{Lorenz,Lorenz1,Uehara} and its cosmological \cite{Barrow,Romero,Romero1,Kolithic,Pandey,Tretyakova:2011ch,Novikov}, cylindrical \cite{delice1,delice2,baykaldeliceciftci} and other \cite{Lee} applications  were discussed in previous works. Here the value of $\Lambda$ in BD$\Lambda$ theory may be different from its value in GR$\Lambda$ theory.  We call the action (\ref{BD-action}) as the BD$\Lambda$ action.  Note that as $\phi$ becomes a constant, this theory reduces to GR$\Lambda$ theory. However, this action is not the only action which reduces to GR$\Lambda$ when $\phi$ is set to a constant. One can replace $2\Lambda \phi$ term with an arbitrary potential term $V(\phi)$  to obtain the following action
\begin{equation}\label{BDaction}
\mathcal{S}_{BDV}=\int \sqrt{|g|}d^4 x \left\{ \frac{1}{16 \pi}\left[\phi R-\frac{\omega}{\phi}g^{\mu\nu}\partial_\mu \phi \partial_\nu \phi -V(\phi)\right] +\mathcal{L}_{matter} \right\}.
\end{equation}
which can be called as the BD action with a potential, i.e., BDV theory. Here, $V(\phi)$ plays the role of a variable cosmological term and  when $\phi$ is set to a constant, the action (\ref{BDaction}) also reduces to GR$\Lambda$ theory. There are various works considering the addition of such a potential term and its various implications to BD theory \cite{Endo,Dereli,Will,Alsing-berti-will,Faraonipot,Faraonibook,Hohmann,Elizalde,Tretyakova}  in various contexts.  In general  BD$\Lambda$ and BD$V$ theories and also all possible BD$V$ theories having different potentials  may have different characteristics and lead to different physics. Hence, there is an arbitrariness  in the generalization of GR$\Lambda$ theory to scalar-tensor theories. One possible method of identifying differences of different gravitation theories is to obtain their weak field solutions and  compare  with the results of GR. Therefore, here  we want to discuss the weak field solutions of the theories (\ref{BD-action}) and (\ref{BDaction}) in the presence of a constant curvature background. Since these different choices may have different characteristics, we have to consider these theories separately, though we try to use a unified treatment as  much as possible in the text. For example,  when we discuss the field equations and their linearization of the actions (\ref{BD-action}) and (\ref{BDaction}) below, to avoid repeated similar equations, we will present those equations for the action (\ref{BDaction}) but keep in mind that for the case (\ref{BD-action}) we have to replace $V(\phi)=2\Lambda\phi$ in the relevant equations. We will present the result of both theories separately whenever their distinction is important.

 The extended BD actions (\ref{BD-action},\ref{BDaction}) we are considering can also be expressed in other frames \cite{Faraonibook,Fujii-maeda}, such as Einstein or string frames, by considering appropriate conformal transformations. For example, the following conformal transformation,
 \begin{eqnarray}
\tilde{g}_{\mu\nu}=\phi\, g_{\mu\nu}, 
 \end{eqnarray}
and the redefinition of the scalar field
\begin{equation}
\tilde\phi=\sqrt{\frac{2\omega+3}{16\pi}}\ln\phi,
\end{equation}
bring the BDV action into the Einstein frame as given by
\begin{eqnarray}
\mathcal{S}_{EV}=\int\left\{\sqrt{|\tilde g|}d^4 x \left[\frac{\tilde R}{16\pi}-\frac{1}{2}\tilde g^{\mu\nu}\partial_\mu \tilde \phi\, \partial_\nu \tilde \phi +\frac{V[\phi(\tilde\phi)]}{[\phi(\tilde\phi)]^2}\right]   +[\phi(\tilde\phi)]^{-2} \mathcal{L}_{matter}\right\}.
\end{eqnarray}
In this frame, the role of the scalar field is changed from being a part of gravitational interaction to a canonical scalar-field matter-energy distribution permeating all points of the spacetime. Moreover, it also couples to the matter Lagrangian nonminimally. Hence, the manifold is no longer Riemannian and the test particles do not follow geodesics. In Jordan frame, however, the scalar field is a part of gravitational interaction. Therefore, the theory is a metric theory and test particles follow geodesics in this frame. Actually, there was a debate on which  of these frames are physical or are they equivalent or not. For a review of this debate, see for example  \cite{FaraoniGunzig}. Here we share the original idea  of transformation of units \cite{BD} of Brans and Dicke which says that both frames are equivalent and give same physical results. After some works concerning this debate, there seems to be a concencus on that all these frames are mathematically  and physically equivalent \cite{FaraoniNadeau,Chiba,Quiros,Postma1,Postma,Rondeau}. Namely, a physical quantity measured on a certain frame does not depend on the chosen frame, if the transformations between frames is properly used.
Hence, the choice of frame is a matter of convenience since some calculations can be more easily performed in a particular frame.  Therefore, in this work, since we will investigate the motion of test particles in the later stages of this paper, we prefer to work in the Jordan frame. This is because this frame has a calculational advantage, since in this frame test particles follow geodesics and we do not want to deal with the fifth force arising from the modifications of geodesics equations in Einstein frame.

The action of a modified gravity theory ($f(R) $ theory) that is very popular in the recent years \cite{modgr1,modgr2,modgr3,modgr4,modgr5} is 
\begin{equation}
\mathcal{S}=\frac{1}{2\kappa}\int \sqrt{|g|} d^4x\left( f(R) +\mathcal{L}_{matter} \right).
\end{equation}
It is well known \cite{modgr1,modgr2,modgr3,modgr4,modgr5} that under a Legendre transformation the $f(R)$ theories become  equivalent to the BDV theory  (\ref{BDaction}) for specific values of $\omega$, namely  $\omega=0$ for metric and $\omega=-3/2$ for Palatini $f(R)$ theories. Thus, BDV theory with arbitrary $\omega$ leads to a more general treatment, includes both $f(R)$ theories as special cases. Therefore, we will consider the theory (\ref{BDaction}) without any restriction on $\omega$ in this paper, except for $\omega=-3/2$ case. In this case, the scalar field becomes an auxillary field with no dynamics and therefore we will not discuss $\omega=-3/2$  case in this paper.  The results for metric $f(R)$ theory can be recovered from BDV theory by setting $\omega=0$ in the resulting expressions.

The Jordan frame field equations of the action (\ref{BDaction}) can be written as
\begin{eqnarray} 
&&G_{\mu\nu}=\frac{8\pi}{\phi}T_{\mu\nu}+\frac{\omega}{\phi^2}\left(\nabla_\mu \phi \nabla_\nu \phi -\frac{1}{2}g_{\mu\nu} \nabla^\alpha \phi \nabla_\alpha\phi \right)+\frac{1}{\phi} \left(\nabla_\mu \nabla_\nu\phi-g_{\mu\nu} \Box_g \phi \right)-\frac{V(\phi)}{2\phi}g_{\mu\nu},\\
&& \Box_g\phi=\frac{1}{2\omega+3}\left(8\pi\, T+\phi \frac{dV(\phi)}{d\phi}-2V(\phi) \right),
\end{eqnarray}
  where here $T$ is the trace of the matter energy-momentum tensor $T_{\mu\nu}$ and $\Box_g$ is the D'Alembertian operator with respect to the full metric. Now, let us consider the weak field expansion of the above field equations. Hence,  the space time metric and the BD scalar field can be expanded as
\begin{eqnarray}\label{weak1}
&&g_{\mu\nu}=\eta_{\mu\nu}+h_{\mu\nu},\quad g^{\mu\nu}=\eta^{\mu\nu}-h^{\mu\nu},\\
&&\phi=\phi_0+\varphi, \nonumber
\end{eqnarray}  
where $\eta_{\mu\nu}=\mbox{diag}(-1,1,1,1)$ is the Minkowski metric, $h_{\mu\nu}$  is the tensor representing small deviation from flatness, $\phi_0$ is a constant value of the scalar field and $\varphi$ is a small perturbation to the scalar field i.e., $|h_{\mu\nu}| \ll 1$  and  $\varphi \ll 1$. Using the above expansion, defining a new tensor \cite{Will}
\begin{equation}\label{weak2}
\theta_{\mu\nu}=h_{\mu\nu}-\frac{1}{2} \eta_{\mu\nu}h-\eta_{\mu\nu} \frac{\varphi}{\phi_0},
\end{equation}
together with the gauge
\begin{equation}\label{gauge-cond}
\theta^{\mu\nu}_{\phantom{aa};\nu}=0,
\end{equation}
the weak field BD field equations, up to \emph{second order}, become \cite{Will}
\begin{eqnarray}\label{weak-second-order}
&&\Box_\eta\theta_{\mu\nu}=-\frac{16 \pi}{\phi_0}\left(T_{\mu\nu}+\tau_{\mu\nu} \right)+\frac{V_{lin}}{\phi_0}g_{\mu\nu},
\\
&&\Box_\eta \varphi=16 \pi S. \label{weak-second-order-sc}
\end{eqnarray} 
Here $\tau_{\mu\nu}$ is the energy-momentum pseudo tensor involving quadratic terms and $\Box_\eta=\eta^{\mu\nu}\partial_\mu \partial_\nu$ is the D'Alembertian of the Minkowski spacetime. The term $S$ is given by
\begin{equation}\label{S}
S=\frac{1}{4\omega+6} \left[ T \left(1-\frac{\theta}{2}-\frac{\varphi}{\phi_0} \right)+\frac{1}{8\pi}\left( \phi \frac{dV}{d\phi}-2V \right)_{lin} \ \right]+\frac{1}{16\pi}\left( \theta^{\mu\nu}\varphi_{,\mu\nu}+\frac{\varphi_{,\nu}\varphi^{,\nu}}{\phi_0}\right).
\end{equation}
 where in deriving $S$  the relation between Minkowski and curved D'Alembertian operators is used
\begin{equation}
\Box_g=\left(1+\frac{\theta}{2}+\frac{\varphi}{\phi_0} \right)\Box_\eta-\theta^{\mu\nu}\varphi_{,\mu\nu}-\frac{\varphi_{,\nu}\varphi^{,\nu}}{\phi_0}+O(\mbox{higher terms}).
\end{equation}
Here subtext $lin$ in equations  (\ref{weak-second-order}) and (\ref{S})  means that these terms must also be  properly linearized. 
In our setting, we want to discuss the case where the space time is not asymptotically flat but asymptotically de Sitter, therefore we need an effective cosmological constant in the linearized field equations. Since  this term can be obtained for either of the  actions (\ref{BD-action}) and (\ref{BDaction}) differently, now we have to discuss these cases separately.

\subsection{Linearised field equations of Brans-Dicke Theory with a cosmological constant (BD$\Lambda$ case) }

For the action (\ref{BD-action}), the terms involving the potential $V$ can be expanded as\begin{eqnarray}
V(\phi)\approx 2\Lambda\, \phi_0,\quad  \left( \phi \frac{dV}{d\phi}-2V \right)\approx - 2\Lambda\, \phi_0,
\end{eqnarray}
where the terms linear in $\Lambda$ or $\varphi$ are kept and the terms of the order $\Lambda\varphi$  or higher  are ignored. Using these, we obtain the following linearised field equations
\begin{eqnarray}\label{weak-first-order-met}
\Box_\eta\theta_{\mu\nu}=-\frac{16 \pi}{\phi_0}\,T_{\mu\nu} +2\Lambda \eta_{\mu\nu},
\\
\Box_\eta\varphi=\frac{8 \pi\, T}{2\omega+3}-\frac{2}{2\omega+3}\Lambda \phi_0.
\label{weak-first-order-sc}
\end{eqnarray} 
In the chosen parametrization (\ref{weak2}) and  gauge (\ref{gauge-cond}), the tensor  equations (\ref{weak-first-order-met}) have similar structure to weak field $GR\Lambda$ equations \cite{Bernabeu}. Thus, all the differences from $GR$, namely the effects of the $BD$ scalar field will be originated from scalar field equation  (\ref{weak-first-order-sc}).   
The first important observation is that for this case the scalar field is \emph{massless}. This means that $BD\Lambda$ theory has the similar structure with $BD$ theory where the scalar field has a long range and the existence of a cosmological constant does not change this behavior. Hence, in $BD\Lambda$ theory the cosmological constant does not change the local behavior of the scalar field and acts as a background curvature similar to $GR$ and responsible for the asymptotical nonflatness. Therefore, $BD\Lambda$ theory is a natural generalization of $GR\Lambda$ theory to BD theory, merging the properties of both theories into a single unified theory.
  \subsection{Linearised field equations of Brans-Dicke Theory with an arbitrary potential (BDV case)}

Now we consider the action (\ref{BDaction}). We suppose that the arbitrary potential $V$ is a   well behaving function of its argument  and it is Taylor expandable around a constant value of the scalar field, namely $\phi=\phi_0$ as:
\begin{equation}
V(\phi)=V(\phi_0)+V'(\phi_0)\varphi+\frac{1}{2}V''(\phi_0)\varphi^2+\ldots
\end{equation}
Here $'$ means partial derivative with respect to the scalar field. In the previous works considering this action \cite{Will,Alsing-berti-will}, asymptotic flatness was assumed, which requires vanishing of the first two terms in the expansion. Here we do not impose such a condition, but it might be reasonable to expect $\phi_0$ to be a minimum of this potential for stability, which require $V'(\phi_0)$ to be vanishing. Then, the relevant terms in the \emph{linearised} field equations can be written as
\begin{equation}
V(\phi)g_{\mu\nu}\approx V(\phi_0)\,\eta_{\mu\nu},\quad \left( \phi \frac{dV}{d\phi}-2V \right)\approx \phi_0 V''(\phi_0)\varphi-2 V(\phi_0)
\end{equation}
 and the field equations  (\ref{weak-second-order},\ref{weak-second-order-sc}),  in the \emph{first order} in $\theta$, $\varphi$  and $V_0$, become
\begin{eqnarray}\label{weak-first-order-pot-met}
&&\Box_\eta\theta_{\mu\nu}=-\frac{16 \pi}{\phi_0}\,T_{\mu\nu} +\frac{V_0}{\phi_0} \eta_{\mu\nu},
\\
&&(\Box_\eta -m_s^2)\varphi=\frac{8 \pi T}{2\omega+3}-\frac{2 V_0}{2\omega+3}.
\label{weak-first-order-pot-sc}
\end{eqnarray} 
 Here we used the abbreviations
 \begin{equation}\label{massterm}
V_0\equiv V(\phi_0), \quad m_s^2\equiv\frac{\phi_0}{2\omega+3}V''(\phi_0)>0 .
 \end{equation} 
In the parametratization  (\ref{weak2}) and gauge (\ref{gauge-cond}) considered, the tensor equation (\ref{weak-first-order-pot-met}) still has the same structure with the weak field equations of GR$\Lambda$ theory \cite{Bernabeu}. The effect of the scalar fied and arbitrary potential is encoded into the scalar field equation (\ref{weak-first-order-pot-sc}). It is clear that the minimum of potential $V_0$ plays the role of a constant curvature background or cosmological constant which is responsible for the asymptotic non-flatness. However, there is a slight difference between $\Lambda$ in BD$\Lambda$ theory and $V_0$ in BDV theory. The terms in the tensor equations of both theories can be made similar by defining $V_0=2\Lambda_0\phi_0$ (we put the subscript $\Lambda_0$ to distinguish these theories). However, even using this redefinition, the coefficients related to cosmological constant $\Lambda$ and $\Lambda_0$  of the scalar field equations (\ref{weak-first-order-sc}) and (\ref{weak-first-order-pot-sc}) will  have different coefficients, $2\Lambda\phi_0/(2\omega+3)$ versus $4\Lambda_0 \phi_0/(2\omega+3)$. This is because the $\Lambda$ term in BD$\Lambda$ theory is linear in scalar field whereas $V_0$ is zeroth order term in the expansion of $V(\phi)$. The effects of this difference will be seen in the solutions of both theories presented in the next section and also at the physical quantities such as advance of perihelion due to these terms.

Another important difference of these theories is that, this theory leads to a \emph{massive} scalar   field \cite{Will,Alsing-berti-will}  with the mass term $m_s$ defined as (\ref{massterm}) proportional to second derivative of the arbitrary potential $V(\phi)$. The effect of the mass term is to make the scalar field short-ranged. As we will see in the next section, solutions representing isolated systems, such as a point particle, contain  Yukawa-like terms, which gives a characteristic range $l_c\sim 1/m_s$ where the scalar field related to this source freezes out  and the physical properties due to this source becomes indistinguishable  from $GR$ outside this range.  This behavior is in contrast to BD theory where the field has a long range. Hence, the introduction of an arbitrary potential changes the range of the scalar field.  Since metric $f(R)$ theory is equivalent to $BDV$ theory for $\omega=0$, this behaviour persists in this theory too. Note that, the behaviour of scalar field  related to minimum of the potential is still long range and persists outside $l_c$. Hence, asymptotically nonflat weak field solutions make it possible to open a new window to test these massive BD theories, otherwise they are indistinguishable from GR outside this range. For example, as we will see in the next section, the advance of perihelion has corrections due to minimum of potential, $V_0$. This correction is negligable  for very light scalar field where $l_c\rightarrow \infty$ and solar system tests require a very large $\omega$ for this case. For a heavy scalar case, however, since $l_c\rightarrow 0$, scalar field due to mass of the isolated source is already frozen and solar system tests have become insensitive to $\omega$. Hence, since $\omega$ can take  arbitrary values, the correction terms involving $\omega$ due to a minimum of potential can be very different from corresponding $GR\Lambda$ solutions.

\section{Solutions to linearized field equations for a point mass}

Having obtained linearized metric and scalar field equations in the chosen gauge for both theories, the next step would be to obtain a physically relevant solution to these theories. Hence, in the following,  we consider a static point mass solution as a source for both  theories. Note that, as far as we know, weak field equations of BD$\Lambda$ theory with nonzero $\Lambda$ for a point particle as a source is not discussed before. For BDV theory however, due to its equivalence with $f(R)$ theory, its weak field solutions derived for a nonvanishing $V_0$ \cite{olmo,olmo1,olmo11} using a slightly different method. The physical applications we will consider of both theories for nonzero $\Lambda$ or $V_0$, namely, advance of perihelion, deflection of light, gravitational redshift, and galactic and inter-galactic dynamics were not discussed before.
    
\subsection{A point mass term as a source for BD$\Lambda$ theory}

We now consider a point particle located at $\bar r=0$, where $\bar r^2=\bar{x}^2+\bar{y}^2+\bar{z}^2$, described by
\begin{equation}\label{Emmass}
T_{\mu\nu}=m \,\delta(\bar r)\,\mbox{diag}(1,0,0,0).
\end{equation}
Then, the scalar field equation (\ref{weak-first-order-sc}) has the solution 
\begin{equation}\label{potentialsoln}
\varphi(\bar r)=\frac{2 m}{(2\omega+3)}\frac{1}{\bar r}-\frac{\Lambda \phi_0}{3(2\omega+3)}\bar r^2.
\end{equation}

The advantage of using the gauge (\ref{weak2}) and (\ref{gauge-cond}) is that the resulting tensor equations are decoupled from scalar  field.  The tensor equation (\ref{weak-first-order-met}) for a diagonal metric ansatze, together with  the gauge condition  (\ref{gauge-cond}) yield the following non vanishing components for the solution
\begin{eqnarray}\label{weaksolntheta}
&&\theta_{00}=\frac{4m}{\phi_0}\frac{1}{\bar r}-\frac{\Lambda}{3}\bar r^2, \quad
\theta_{xx}=\frac{\Lambda }{2} (y^2+z^2),\quad \\
&& \theta_{yy}=\frac{\Lambda }{2} (x^2+z^2),\quad  \theta_{zz}=\frac{\Lambda}{2} (x^2+y^2).\nonumber
\end{eqnarray}
Using the trace of $\theta=\theta_{\mu}^{\mu}$ given by
\begin{equation}
\theta=-\frac{4m}{\phi_0\, \bar r}+\frac{4\Lambda}{3}\bar r^2,
\end{equation}
and the inverse of (\ref{weak2}), the nonzero components  of the metric perturbation term becomes
\begin{eqnarray}
&&h_{00}=\frac{2m}{\phi_0 \bar r}+\frac{\Lambda}{3} \bar r^2+\frac{\varphi}{\phi_0},\quad \nonumber \\
&&h_{ij}=\left[\frac{2m}{\phi_0 \bar r}-\frac{ \Lambda}{6}(\bar r^2 + 3\, x_i^2)-\frac{\varphi}{\phi_0}\right]\delta_{ij},\quad 
(i,j=1,2,3).  \label{metricnonisotropic}
\end{eqnarray} 
Clearly, the effects of the nonminimally coupled scalar field reveal themselves as the last terms in the metric perturbation tensor. The presence of this nontrivial scalar field may have some physical consequences  such as it can modify  test particle trajectories compared to corresponding GR results  discussed, for example, in \cite{Bernabeu}.

The solution presented above in equation (\ref{metricnonisotropic}) is not in  isotropic coordinates. To express this solution in isotropic coordinates we may consider the following coordinate transformations \cite{Bernabeu}
\begin{equation}\label{transtoisotr}
\bar x^i={x'^i}+\frac{\Lambda}{12}{x'^i}^3.
\end{equation}
Under these transformations (\ref{transtoisotr}) from barred to primed coordinates, the metric perturbation terms, up to linear order in $M$ and $\Lambda$  become
\begin{eqnarray}
&&h'_{00}=\frac{2m}{\phi_0 r'}+\frac{\Lambda}{3}r'^2+\frac{\varphi'}{\phi_0},%\nonumber 
\\
&&h'_{ij}=\left(\frac{2m}{\phi_0 r'}-\frac{\Lambda}{6}r'^2-\frac{\varphi'}{\phi_0} \right)\delta_{ij},
\\
&&\varphi'=\frac{2 m}{(2\omega+3)}\frac{1}{r'}-\frac{\Lambda \phi_0}{3(2\omega+3)}r'^2. \label{varphi}
\end{eqnarray}

The explicit expressions of the field variables,  with the help of (\ref{varphi}),   can be expressed  as
\begin{eqnarray}
&&g'_{00}=-1+\frac{2m}{\phi_0 r'}\left(1+\frac{1}{2\omega+3} \right)+\frac{\Lambda r'^2}{3} \left(1-\frac{1}{2\omega+3}\right),\label{g00}\\
&&g'_{ij}=\left[ 1+\frac{2m}{\phi_0 r'}\left(1-\frac{1}{2\omega+3}\right) -\frac{\Lambda r'^2}{6}\left(1-\frac{2}{2\omega+3} \right) \right]\delta_{ij},\label{gij}\\
&&\phi'=\phi_0\left( 1+\frac{2m}{(2\omega+3)\phi_0 r'}- \frac{\Lambda r'^2}{3(2\omega+3)} \right).\label{phii}
\end{eqnarray} 
As well known \cite{BD}, the mass term in $g_{00}$ must be related  with weak field GR or Newton potential of a point mass, then  $\phi_0$ must be equal to
\begin{equation}\label{phi01}
\phi_0=\frac{2\omega+4}{2\omega+3}.
\end{equation}
This implies that, since it is defined for an asymptotically flat space time, when $\Lambda=0$, the post-Newtonian parameter $\gamma$
 is
\begin{equation}\label{gammabd}
\gamma_{BD}=\frac{h_{ij}|_{i_=j}}{h_{00}}=\frac{\omega+1}{\omega+2}.
\end{equation}
This result, together with the observational result of Cassini mission \cite{cassini}, i.e. $\gamma_{observed}-1=(2.1\pm 2.3)10^{-5}$ which sets $\gamma\sim 1$, implies that BD parameter must satisfy $\omega>40.000$ for BD theory. For these large value of $\omega$, the above solution given by (\ref{g00}) and (\ref{gij}) becomes indistinguishable from the corresponding $GR\Lambda$ solution \cite{Bernabeu}. 

 Note that, in the case of vanishing $\Lambda$, the metric (\ref{g00}) and (\ref{gij}) and scalar field (\ref{phii})  reduce to the linearized BD solution \cite{BD,Brans}. These solutions reduce to corresponding linearized GR$\Lambda$ solutions presented in \cite{Bernabeu} in the limit ($\omega\rightarrow \infty,\quad \phi_0\rightarrow 1$). Hence the above solution has correct limits. Note that the limit $\omega\rightarrow \infty$ does not always reduce the theory to GR. For a further discusson of the GR limit of  BD theory, see, for example \cite{limit1,Romero1,limit3,limit4,limit5,limit6} and references therein.

\subsection{A point mass term in the BDV theory}

For a point mass  $m$ given in (\ref{Emmass}) the corresponding solution of the scalar field equation (\ref{weak-first-order-pot-sc}) reads: 
\begin{eqnarray}\label{potmassive}
&&\varphi(\bar r)=\frac{2 m}{(2\omega+3)}\frac{e^{-m_s \bar r}}{\bar r}-\frac{V_0}{3(2\omega+3)}\bar r^2.
\end{eqnarray}
Note that if we set 
\begin{equation}\label{lambdaV0}
\frac{V_0}{2\phi_0}=\Lambda_0,
\end{equation}
 then the first order metric equation (\ref{weak-first-order-pot-met}) becomes  exactly the same as (\ref{weak-first-order-met}) when $\Lambda$ is replaced by $\Lambda_0$. Thus, we can directly use the solutions (\ref{weaksolntheta}) and also for the metric perturbation terms $h_{\mu\nu}$ given in (\ref{metricnonisotropic}) in this case as well.  Moreover, with this choice (\ref{lambdaV0}),  we can use exactly the same transformations (\ref{transtoisotr}) to bring the metric solution into isotropic coordinates.   With this notation,  the differences between both theories  are encoded in $\varphi'$ term, which has the same form  with (\ref{potmassive})  for this case with $\bar r$ replaced with $r'$.
 Then, full metric and scalar field of weak field equation becomes
 \begin{eqnarray}
&&g'_{00}=-1+\frac{2m}{\phi_0 r'}\left(1+\frac{e^{-m_s \bar r}}{2\omega+3} \right)+\frac{V_0\, r'^2}{6\phi_0} \left(1-\frac{2}{2\omega+3}\right),\label{g00V}\\
&&g'_{ij}=\delta_{ij}\left[ 1+\frac{2m}{\phi_0\, r'}\left(1-\frac{e^{-m_s \bar r}}{2\omega+3}\right) -\frac{V_0\, r'^2}{12\, \phi_0}\left(1-\frac{4}{2\omega+3} \right) \right],\label{gijV}\\
&&\phi'=\phi_0\left( 1+\frac{2m\, e^{-m_s \bar r}}{(2\omega+3)\phi_0\, r'}- \frac{V_0\, r'^2}{3\,\phi_0\,(2\omega+3)} \right).
\end{eqnarray} 
Note that this solution is discussed before by using a slightly different approach \cite{olmo,olmo1,olmo11}. The vanishing $V_0$ case is known as massive BD theory and its weak field solutions were presented before \cite{Will,Steinhardt-will,Alsing-berti-will}. Also for vanishing $V_0$, the post-Newtonian parameter $\gamma$   
 becomes position dependent \cite{olmo,olmo1,olmo11,Perivolaropoulos}:
\begin{equation}
\gamma(r)=\frac{1-\frac{e^{-m_s r}}{2\omega+3}}{1+\frac{e^{-m_s r}}{2\omega+3}}.
\end{equation} 

 The observational result of Cassini mission, namely $\gamma_{observed}-1=(2.1\pm 2.3)10^{-5}$, which sets $\gamma(r)\sim 1$, can be applied to $\gamma(r)$ in several ways. The first one is to set $e^{-m_s r}\rightarrow 0$, which requires $m_s\rightarrow \infty$, namely the mass of the scalar field must be very heavy. For this case $\gamma=1$ irrespective of the value of $\omega$. If this is not the case, i.e.  if the mass is not large and if $e^{-m_s r}$ is $O(1)$ which requires very light scalar mass as $m_s\rightarrow 0$, then $\gamma(r)$ reduces to $\gamma_{BD}$ given in equation (\ref{gammabd}) and the limit $\omega>40.000$ is again emerged. For intermediate values of $m_s$,  however, a numerical investigation is required to find the observationally allowed regions of the parameter space $(m_s,\omega)$ and this analysis is done in the work \cite{Perivolaropoulos}.  
 
  Let us now discuss the effective gravitational constant $\phi_0$ for these theories. In the vanishing of the $V_0$ term the value of $\phi_0$ is fixed by the requirement that  the theory must have a correct Newtonian limit, which requires the investigation of the term containing the mass of the source in $g_{00}$. The exponential term spoils this expression but for very light or heavy scalar field mass cases $\phi_0$ can be fixed as discussed below \cite{Will,Perivolaropoulos,Alsing-berti-will}:
\begin{itemize}
\item{} For a very massive potential, i.e. $m_s\gg 1$, we can ignore the terms with exponential factor and can set $\phi_0=1.$ 

  \item{} For a very light scalar mass case, i.e.  $m_s\ll 1$, we can expand $e^{-m_s /r}$ term in $g_{00}$ in series,  keep the first term and compare with the Newtonian potential of a point mass. This procedure  yields that we must have $\phi_0=(2\omega+4)/(2\omega+3)$, as in the original BD theory \cite{BD}.
\end{itemize} 
For the intermediate values of $m_s$, the above prescription does not work to fix $\phi_0$, but one can define an effective gravitational coupling term  involving the exponential term as
\begin{equation}
G(r)=\left(1+\frac{e^{-m_s r}}{2\omega+3} \right)\frac{1}{\phi_0}.
\end{equation}
 A numerical investigation of such case is given in \cite{Perivolaropoulos}   by using the powerful PPN approach, which requires an asymptotically flat spacetime so that $V_0$ must be vanishing. In our work we will not discuss such an investigation since we want to discuss the case where the spacetime is not asymptotically flat. 

As we have discussed in the previous section, another difference between the weak field solutions of BD$\Lambda$ and BDV theories is that the factors involving $\omega$ in the metric and scalar field expressions of  $\Lambda$  or $V_0$ have some differences. The result reflects the fact that their couplings with the scalar field are different. Namely $V_0$ is constant whereas the term involving  $\Lambda$ is linear in the scalar field. 

\subsection{Solutions in Schwarzschild Coordinates}

Here we will bring both of the solutions given in subsections (III-A) and (III-B) to Schwarzschild type coordinates.  
This can be done by the following transformations
    and definitions
\begin{eqnarray}\label{BDCCSch}
r'=r\left(1-\frac{m}{\phi_0 r}+\frac{1}{12}\Lambda r^2+\frac{\varphi}{2\phi_0}  \right),
\end{eqnarray}
with the result
\begin{equation}\label{Schmetric}
ds^2=-\left(1-\frac{2m}{\phi_0 r}-\frac{\Lambda r^2}{3}-\frac{\varphi}{\phi_0} \right)dt^2+\left( 1+\frac{2m}{\phi_0 r} +\frac{\Lambda r^2}{3}+\frac{\alpha \, r }{\phi_0} \right)d r^2+r^2 d\Omega_2^2.
\end{equation}
Here $\varphi$ is given by
\begin{equation}\label{varphiBDL}
 \varphi=\frac{2 m}{(2\omega+3)}\frac{1}{r}-\frac{\Lambda \phi_0}{3(2\omega+3)}r^2,
\end{equation}
for BD$\Lambda$ theory and
\begin{equation}\label{varphiBDV}
 \varphi= \frac{2m\, e^{-m_s  r}}{(2\omega+3)\, r}- \frac{V_0\, r^2}{3\,(2\omega+3)},
\end{equation}
for BDV theory. The function $\alpha(r)$ in (\ref{Schmetric}) is defined as
\begin{equation}
 \alpha=\frac{d  \varphi}{d r}. \label{alpha}
\end{equation}
This form of the metric  (\ref{Schmetric}) is suitable to represent both solutions with the same metric. Difference from corresponding GR$\Lambda$ solution is encoded in $\phi_0$, $\varphi$   and $\alpha$. 
 \section{Motion Of Test Particles}
 
Now we will discuss  the effects of the  point mass and the presence of the cosmological term  on the motions of test particles and photons. In order to discuss these effects and compare with the previous results exist on the literature, we choose to work in the Schwarzschild like coordinates, which can be written as
\begin{equation}\label{Schwarzschild}
ds^2=-A(r)\,dt^2+B(r)\, dr^2+r^2\left(d\theta^2+\sin^2\theta\, d\Phi^2 \right).
\end{equation}
 We also consider equatorial motion by setting $\theta=\pi/2$, then the Lagrangian of the test particles or photons can be written as 
\begin{equation}
   2\mathcal{L}=-A\, \dot{t}^2+B\, \dot{r}^2+r^2\, \dot{\Phi}^2.
   \end{equation}
Here overdot means derivative with respect to proper time for time-like particles and an affine parameter for photons.  Symmetries of this space time results two first integrals of motion, given by
\begin{equation}\label{firstintegrals}
\dot{t}=-\frac{E}{A}, \quad \dot{\Phi}=\frac{L}{r^2}.
\end{equation}   
Here $E$ and $L$ are related with specific energy and angular momentum of test particles. 
   We can use these results into the metric itself to obtain a radial equation of motion
   \begin{equation}\label{radialeom}
   \dot{r}^2=\frac{1}{B}\left(\varepsilon+\frac{E^2}{A}-\frac{L^2}{ r^2} \right),
   \end{equation}
   where $\varepsilon=0$ for photons and $\varepsilon=-1$ for timelike particles. We can obtain an equation for orbit by dividing this expression by $\dot{\Phi}^2$ with the result
   \begin{equation}\label{orbiteqn}
\left(\frac{dr}{d\Phi}\right)^2=\frac{\dot{r}^2}{\dot{\Phi}^2}=r^4\left(\frac{\varepsilon}{L^2 B}+\frac{E^2}{L^2 A B}- \frac{1}{Br^2}\right).
\end{equation}
Hereafter we may  analyze the equations (\ref{radialeom}) or (\ref{orbiteqn}) for different types of motion for the  two different solutions we have obtained. We should also read the appropriate metric functions   $A$ and $B$  (\ref{Schwarzschild}) from (\ref{Schmetric}) for the solutions we are discussing and keep the terms in the linear order in mass and cosmological terms in the expressions.   It is customary to use  inverse radial coordinate defined by
\begin{equation}\label{urtrans}
u=\frac{1}{r}
\end{equation}
  to obtain a modified Binet equation and solve the resulting equation. Then, using the transformation (\ref{urtrans}), the resulting equation can be written in the linearized order as 
 \begin{eqnarray}
 \left(\frac{du}{d\Phi}\right)^2+u^2&&=\left(\frac{E^2+\varepsilon}{L^2}\right) +\frac{\Lambda}{3} - \frac{2\varepsilon \,m \,u}{\phi_0 L^2} + \frac{2m  u^3}{\phi_0} - \frac{\varepsilon\Lambda}{3L^2u^2}+\frac{E^2 \varphi(u)}{L^2} - \left(E^2+\varepsilon-L^2 u^2\right)\frac{\alpha(u)}{L^2\, u\,\phi_0}.
\end{eqnarray}
By differentiating this equation with respect to $\Phi$, one obtains a modified Binet equation as
 \begin{equation}\label{BDBinet}
\frac{d^2u}{d\Phi^2}+u=-\frac{\varepsilon\, m }{\phi_0 L^2}+ \frac{3 m u^2}{\phi_0}+ \frac{\varepsilon \, \Lambda }{3 L^2 u^3 } +\frac{1}{2}\frac{\partial}{\partial u} \left[\frac{E^2 \varphi(u)}{L^2} - \left(E^2+\varepsilon-L^2 u^2\right)\frac{\alpha(u)}{L^2\, u\,\phi_0} \right].
\end{equation}
Hence, we have obtained a modified Binet equation  which is useful for both BD$\Lambda$ and BDV theories. One can further analyze this equation  by appropriately choosing  the values of the functions $\varphi, \alpha$    and constant $\varepsilon$. 

\section{Precession of the perihelion of the planets}

In order to derive advance of perihelion for these theories, we may try to use the usual perturbative approach to solve equation (\ref{BDBinet}) for time like particles $\varepsilon=-1,$ together with appropriate values for $\varphi$ and $\alpha$. For example, for BD$\Lambda$ theory, we obtain the following differential equation
\begin{equation}\label{BDLBinet}
\frac{d^2u}{d\Phi^2}+u=\frac{2m( E^2+\omega+1)}{(2\omega+3)\phi_0 L^2}+ \frac{6 m (\omega+1)u^2}{(2\omega+3)\phi_0}- \frac{(E^2+2\omega+1)\Lambda}{3(2\omega+3) L^2 u^3}.
\end{equation}
One can consider the Newtonian elliptical solution as a zeroth order solution of the equation (\ref{BDLBinet}).
However, this approach needs a perturbation extension of geodesic equation (\ref{BDLBinet})  \emph{second} order in mass $m$. But, this is beyond our linearized approximation. Hence, we cannot use the perturbation approach to derive the perihelion advance for both of the theories. But, there are alternative methods and we will use one of them to derive perihelion shift terms due to mass of the source and cosmological terms.

   \subsection{Review of calculation of advance of perihelion by integration of perturbation potential%\label{appendix}
   }

The calculation is based on the principle that in the weak field of a gravitation theory, we may have a potential which has usual Newtonian term for a central mass distribution as well as other correction terms originated from the theories modifying Newtonian theory. Using different methods such as directly integrating geodesics equations \cite{adkins}, precession of a cousin of Runge-Lenz vector, i.e., the Hamilton vector in the modified potential background \cite{chashchina}, or using a modification of Landau-Lifshitz method \cite{chashchina-1}  one obtains perihelion shift $\Delta$ as a one dimensional integral  of the form
\begin{equation}
\Delta=\frac{-2 L}{m e^2}\int_{-1}^{1} \frac{z\, dz}{\sqrt{1-z^2}} \frac{d\tilde{V}(z)}{dz}, \label{adkins-integral}
\end{equation}
 where the coordinate transformation $r=L/(1+e\,z)$ is employed. Here $e$ is eccentricity of elliptic motion. For details of the derivation of this integral we refer to the works \cite{adkins,arakida-perihelion}.
 This perihelion shift implies that the orbit equation has the following form
 \begin{equation}
  u=1/r=\frac{m}{L^2}\left[1+e \cos{(1-\epsilon)\Phi}  \right], \quad \Delta=2\pi \epsilon.
 \end{equation}

 In equation (\ref{adkins-integral}), $\tilde V(z)$ contains the modification terms to the Newtonian potential for central motion which can be read from the full potential of the gravity theory considered of a point mass as given by
 \begin{equation}
U(r)=-\frac{m}{r}+\frac{L^2}{2r^2}+\tilde{V}(r). 
 \end{equation}
 So here we just need to find the $\tilde V(r)$ term and replace into the integral (\ref{adkins-integral}). To do this we consider the radial geodesic equation (\ref{radialeom}), which can be written as
 \begin{equation}
  \frac{\dot{r}^2}{2}+U(r)=\tilde E
 \end{equation}
where $\tilde{E}=E^2/2$, and
\begin{equation}\label{U}
 U(r)=\frac{1}{2}\left(1+\frac{L^2}{r^2} \right)\left(1-\frac{2m}{\phi_0 r}-\frac{\Lambda r^2}{3}-\frac{\alpha r}{\phi_0}  \right)+\left( \alpha r-\varphi\right)\frac{E^2}{2\,\phi_0}.
\end{equation}
 Hence, we need to evaluate $U(r)$ for the solutions we have found to read $\tilde V(r)$. We will do this in the following subsections for both of the theories we consider.
 
 \subsection{Advance of perihelion for BD$\Lambda$ theory}
 
 Using (\ref{Schmetric}), replacing the value of $\phi_0$  given in (\ref{phi01}) and the fact that for a nonrelativistic motion energy per unit mass has the value $E=1$ \cite{schiffer}, we find that
\begin{equation} 
 \tilde V(r)=-\left(\frac{\omega+1}{\omega+2}\right)\frac{m L^2}{r^3}-\left(\frac{2\omega+2}{2\omega+3}\right)\frac{\Lambda r^2}{6}.
 \end{equation}
 Here, some constant terms are discarded because they do not affect the advance of perihelion. Since the result of integral (\ref{adkins-integral}) is calculated for power-law potentials \cite{adkins,arakida-perihelion}, from their result, the only difference is factors involving $\omega$ in $\tilde V(r)$, so we see that 
 the perihelion shift can be written as
 \begin{equation}\label{perihelionBDLambda}
\Delta_{BD\Lambda}=\frac{\omega+1}{\omega+2}\Delta_E+\frac{2\omega+2}{2\omega+3}  \Delta_\Lambda, 
 \end{equation} 
  where $\Delta_E$ is the usual Einstein value of perihelion shift due to mass  \cite{Einstein} and $\Delta_\Lambda$ \cite{adkins,Kagramanova,arakida-perihelion,Kerr} is GR$\Lambda$ perihelion shift due to cosmological constant. Their expressions are given by
  \begin{eqnarray}
\Delta_{E}
=\frac{6\pi\, m}{a(1-e^2)},\label{DeltaE}\\
\Delta_\Lambda=\frac{\pi\, \Lambda}{m} a^3 \sqrt{1-e^2},\label{DeltaL}
\end{eqnarray}
 and $\Delta_\Lambda$ agrees with the one found in \cite{Rindler} only for $e\rightarrow 0$ as discussed in \cite{arakida-perihelion}. The perihelion shifts due to mass of the source and the cosmological constant of $BD\Lambda$ theory have similar structures with corresponding GR$\Lambda$ theory with same multiplicative factors involving $\omega$. The discussion of some observational consequences  of these results will be  in the Section(\ref{observability}).

 \subsection{Advance of perihelion for BDV theory}
 
 For this case, the potential $U$  given by equation (\ref{U}) becomes
 \begin{eqnarray}
  U(r)&=&\frac{1}{2}\left(1+\frac{L^2}{r^2} \right)\left[1-\frac{2m}{\phi_0 r}-\frac{\Lambda r^2}{3}-\frac{2m}{(2\omega+3)\phi_0} \left( m_s+\frac{1}{r}\right)e^{-m_s r}-\frac{2 V_0 r^2}{3\phi_0}  \right] \nonumber \\
  && -\frac{E^2 m e^{-m_s r}}{(2\omega+3)\phi_0}\left(m_s-\frac{2}{r} \right)+\frac{V_0 E^2 r^2}{6(2\omega+3)\phi_0}.
 \end{eqnarray}
This expression  is complicated and resulting $\tilde V(r)$ which involve terms containing  $e^{-m_s r}$ factor cannot be integrated \cite{adkins} to obtain analytical results. However,  for the following special cases it is possible to obtain analytic results for the advance of perihelion.
 
 \begin{itemize}
  \item{} For a very heavy scalar field, since as $m_s \rightarrow \infty,$ $e^{-m_s r} \rightarrow 0$, the perturbation potential becomes
  \begin{equation}
  \tilde V(r)_{m_s -> \infty}=-\frac{m L^2}{r^3}-\left(\frac{2\omega+1}{2\omega+3} \right)\frac{\Lambda_0 r^2}{6}.
  \end{equation}
From this potential, since $\phi_0=1$  for a heavy scalar field, the advance of perihelion is calculated as
\begin{eqnarray}\label{Perihelionheavy}
 \Delta_{BDV-heavy}=\Delta_E+\left(\frac{2\omega+1}{2\omega+3} \right) \Delta_{\Lambda_0},
\end{eqnarray}
where $\Delta_E$ is the Einstein value (\ref{DeltaE}) for perihelion shift and $\Delta_{\Lambda_0}$ is  the GR$\Lambda$ value (\ref{DeltaL}) with $\Lambda$   is replaced with $\Lambda_0$. Hence, for a very heavy scalar, when the minimum of the potential is zero, then the perihelion shift is indistinguishable from GR value and independent of $\omega$. This is a well-known result that when the scalar field becomes very short range field, weak field tests yield the same results with  GR. However, when the minimum of the potential $V_0$ is not zero, then the resulting perihelion shift has a term due to the minimum of potential having a factor involving $\omega$. If the value of $V_0$ or $\Lambda_0$ would be fixed by a future  observation, then one could put on bounds on $\omega$ even for very massive BD theory. For example for  metric $f(R)$ case $\omega=0$, the perihelion shift due to mass is the same with GR whereas the corresponding term due to the cosmological term is $1/3$ of its  GR$\Lambda$ value with $\Lambda$ is replaced with $\Lambda_0$.

\item{} For a very light scalar field, as $m_s\rightarrow 0$ we can expand $e^{-m_s r}$ in a series and since $m_s$ and $m$ are small we can ignore  the terms such as $m\times m_s$, and using the value of $\phi_0$ given in (\ref{phi01}) for this case, then the perturbation term becomes
\begin{equation}
\tilde V(r)_{m_s\rightarrow 0}=-\left(\frac{\omega+1}{\omega+2}\right)\frac{m L^2}{r^3}-\left(\frac{2\omega+1}{2\omega+3} \right)\frac{\Lambda_0 r^2}{6}.
\end{equation}
This implies the perihelion shift as 
 \begin{eqnarray}\label{Perihelionlight}
 \Delta_{BDV-light}=\frac{\omega+1}{\omega+2}\Delta_E+\left(\frac{2\omega+1}{2\omega+3} \right) \Delta_{\Lambda_0},
\end{eqnarray}
  Hence, for the case when the mass of the scalar field  is very light, the scalar field becomes a long range one, similar to original BD scalar. Thus, the advance of perihelion term of light BD$V$ theory of a point mass becomes exactly the same as the result of BD theory given in \cite{BD}.  For both heavy or light BDV theories, the effect of minimum of the potential has same $\omega$ dependent factor, which is slightly different from the factor of the result of BD$\Lambda$  theory given in equation (\ref{perihelionBDLambda}). For $f(R)$ theory with very light mass,  the perihelion shift due to mass of the source becomes one half of corresponding $GR$ value and  corresponding term due to the minimum of the potential is $1/3$  of the similar term due to cosmological constant for the GR$\Lambda$ solution. Hence a very light scalar field cannot be compatible with solar system tests but one can circumvent this result with some ideas such as chameleon mechanism \cite{Veltman1,Veltman2,modgr4}. We will discuss  some other observational consequences for both light or heavy scalar field mass of the BDV theory in the following subsection.
  \end{itemize}
  
 \subsection{Observability\label{observability}}
We have obtained advance of perihelion for both $BD\Lambda$ theory given in (\ref{perihelionBDLambda})  and for heavy or light $BDV$ theories given in equations (\ref{Perihelionheavy}) and (\ref{Perihelionlight}), due to mass of the source and cosmological constant or minimum of potential, respectively. These expressions have a similar structure to the corresponding GR ones given in (\ref{DeltaE}) and (\ref{DeltaL}). As we have discussed before,  the difference is the different numerical factors involving $\omega$ multiplying these terms. The multiplicative factors due to mass are the PPN parameters $\gamma$  of these theories. For $BD\Lambda$ theory and light $BDV$ theory these parameters are the same as in the original $BD$ theory as given in (\ref{gammabd}). For these cases, the results agree with corresponding GR results for large $\omega$ since for these cases we have  $\omega>40.000$. For a very massive scalar mass case, however, this term is independent of $\omega$ and equal to $GR$ value,  1.  Hence for heavy $BDV$ theory, solar system tests will be satisfied for any value of $\omega$ except $\omega=-3/2$.

\begin{figure}[h]
\begin{centering}
\includegraphics[draft=false,width=7cm]{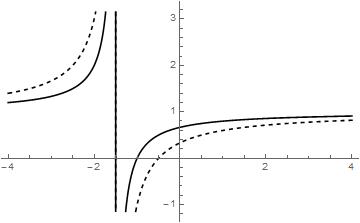}
\caption{The behavior of some factors involving $\omega$. The continuous line represents the ratio $(2\omega+2)/(2\omega+3)$ whereas the dashed line represents $(2\omega+1)/(2\omega+3)$.} 
\label{fig1}
\end{centering}
\end{figure}

When we take into account the effects due to the cosmological constant or minimum of the potential, we see that  these terms have a similar structure to the corresponding $GR\Lambda$  theory. The differences are the existence of multiplicative terms involving $\omega$.  The behaviour of these factors can be seen from figure (\ref{fig1}). As it is clear from this graph, for positive values of $\omega$,  these numerical factors are in the intervals $[2/3,1)$ for $BD\Lambda$ theory  and $[1/3,1)$ for $BDV$ theories for $0\le \omega <\infty$. Therefore, in these intervals, the correction factors cannot make significant order of magnitude changes to these terms. For negative values of $\omega,$ however, these factors may have significant effects, as seen from the graph.  These factors even vanish  at $\omega=-1$ for $BD\Lambda$ and $\omega=-1/2$ for  $BDV$ theories or take negative values for $-2/3<\omega<-1$ for $BD\Lambda$ and $-2/3<\omega<-1/2$  for $BDV$ theories. These factors take positive values for both theories for $\omega<-3/2$. They blow up as seen from graph as $\omega\rightarrow -3/2$. Therefore,  for small and negative values of $\omega$, the deviation from GR can be observed  for these theories, in principle.   

 In sumary, the result of the Cassini experiment  sets a lower bound for BD parameter $\omega$ as $\omega>40.000$ for BD theory and this behavior is also valid for $BD\Lambda$ and  light $BDV$ theories. Hence, for both $BD\Lambda$ and light $BDV$ theories, the multiplying factors of $\omega$ for the terms contributing to the perihelion precession due to the mass of the source and cosmological or minimum potential terms approaches to one as $\omega$ approaches to  $40.000$, as can be seen from figure (\ref{fig1}). Thus, these terms cannot have an effect on the advance of perihelion.  It is also known that, at solar system scales, the effect of  cosmological constant is too small to be observable \cite{Kagramanova,Iorio}. Conversely, one can put lower bounds on $\Lambda$ or $\Lambda_0$  using the results of  \cite{Kagramanova,Iorio}, which is, $\Lambda \le 10^{-41} m^{-2}$ or the same limit for $\Lambda_0$. For heavy $BDV$ theory, however, the local behaviour does not fix $\omega$ and in principle this term can take any value. Most significant effects of the multiplicative factor is at the negative values of $\omega$. Namely, for negative values of $\omega$ there are regions where the factor $(2\omega+1)/(2\omega+3)$ becomes zero, negative, or takes unbounded negative or positive values as $\omega\rightarrow -3/2$ from left or right.  For example, as $\omega\rightarrow -3/2$ a very small minimum potential term $\Lambda_0$, smaller than current observed value of $\Lambda$,  can be compatible with observations. Or conversaly, if $\omega\rightarrow 1/2$, as this factor approaches to zero, a very large minimum potential compared to observed cosmological constant, may be compatible with  observations on perihelion precession. As a result, there may be an observational window to test heavy $BDV$ theories with solar system tests  if for example the contribution of  advance of perihelion due to cosmological constant can be measured with enough sensitivity in the future observations. 
 
\section{Deflection of light rays}

In this part we will discuss deflection of light rays for both BD$\Lambda$ and BDV theories using the geodesic equations derived in section (IV). The effect of cosmological constant on the deflection angle was a topic with opposing views with works confirming \cite{,Rindler-ishak,Ishak-rindler,Ishak-rindler-dosset,light1,light2,light3,light4,light5,light6,light7,light8,light9,Lebedev-lake,Lebedev-lake1,arakida1,light10} or denying \cite{Islam,light11,light12,light13,light14,light15,light16} this effect. Therefore,  here we first give a short summary of this topic in the discussion below. Then, we will focus on such effects for the theories we are considering.

\subsubsection{Calculation of Deflection angle for GR$\Lambda$ theory using Rindler-Ishak method}
Here we review  deflection of light rays from a compact object in GR$\Lambda$ theory in the linear approximation. This requires geodesics  of photons in the corresponding space-time. We will again use Schwarzschild type coordinates (\ref{Schmetric}) for this discussion as well, hence we can use the  orbit equation (\ref{BDBinet}) for photons $\varepsilon=0$.  Note that, whether the  cosmological constant affects the light deflection angle has became a  source of debate and a lot of work is devoted to clarify this issue using different techniques. The reason for this is the fact that, for GR$\Lambda$ theory, the geodesic equation for photons (\ref{BDBinet}) can be reduced  to 
\begin{equation}
\frac{d^2u}{d\Phi^2}+u= 3 m u^2.
\end{equation}
The fact that this equation is independent of cosmological constant term lead to the conclusion \cite{Islam} that cosmological constant has no effect in the light deflection. This is because the  solution of this equation, given by \cite{Rindler}
\begin{equation}\label{light-gr-soln}
u(\Phi)\equiv u_{GR}(\Phi)=\frac{1}{r}  = \frac{\sin \Phi}{R}+\frac{3}{2}\frac{m }{R^2}\left(1+\frac{1}{3}\cos{2\Phi} \right)
\end{equation}
does not involve cosmological constant explicitly, hence orbit is independent of $\Lambda$. Note that the relation between integration constant $R$  and the closest approach distance $r_0$, given by setting $\Phi=\pi/2$ in  (\ref{light-gr-soln}), is
\begin{equation}\label{gr-rel-intcost}
\frac{1}{r_0}=\frac{1}{R}+\frac{m}{R^2}.
\end{equation} 
This  means that in the  orbit equation (\ref{light-gr-soln}) we can replace $R$ with $r_0$ in the linearised approximation. Here $r_0$ is the solution of the equation $dr/d\Phi=0$ and from (\ref{light-gr-soln})  it is given by
\begin{equation}\label{gr-rel-b}
\frac{1}{b^2}+\frac{\Lambda}{3}=\frac{1}{r_0^2}-\frac{2m}{r_0^3}.
\end{equation}
 This means that we can express integration constant $R$  in terms of either $r_0$ or $b$  and $\Lambda$ and the latter choice produce a $\Lambda$ dependence in the bending angle expressions. For an asymptotically flat spacetime, the solution (\ref{light-gr-soln}) implies half bending angle for Schwarzschild spacetime, as $r\rightarrow \infty$,
\begin{equation}
\alpha_E=\frac{2m}{R}=\frac{2m}{b}
\end{equation} 
 where the last equality is valid in the linearised order only. Note that the asymptote $r \rightarrow \infty$ is not valid for Schwarzschild-de Sitter space time because this space time is not asymptotically flat.  One might attempt to obtain a $\Lambda$ dependence by using the relations (\ref{gr-rel-intcost}) and (\ref{gr-rel-b}).   However, this was criticized in \cite{Lebedev-lake} and argued that despite this dependence, the orbit is not affected by $\Lambda$.

  In a pionering work proposed in \cite{Rindler-ishak}, if one considers the measurement of angles which depens on both the local and global geometry of the space-time, bending angle can be shown to be affected by the cosmological constant as well. It turns out that, this problem depends both on how to define and  measure bending angle and also how to specify physical parameters such as impact parameter. However, this is not a generic conclusion and  both depends exactly to the setup used to perform observation to measure, and also the initial conditions as well. Since the spacetime obtained  is not asymptotically flat, the above measurements and definitions will be different from Schwarzschild case and may lack a universal understanding. We refer the latest works for a more complete review of this topic \cite{Ishak-rindler,Lebedev-lake}, and in the latter part of the paper we consider bending of light phenomena for BD$\Lambda$ and BDV theories.  There are many different approaches to this problem but here we  only consider Rindler-Ishak method presented in \cite{Rindler-ishak,Ishak-rindler} in this work. Now, let us review their method and results here.

\begin{figure}[h]
\begin{centering}
\includegraphics[draft=false,width=7cm]{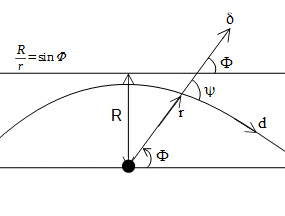}
\caption{The plane graph  corresponding to the orbit  equation given in equation (\ref{light-gr-soln}). 
The one-sided deflection angle is given by $\alpha=\psi-\Phi$ (The figure is adapted from \cite{Rindler-ishak}).}
\label{f1}
\end{centering}
\end{figure}

 Now, consider the case where both source and observer are static. The cosine of the angle between two coordinate directions $d$ and $\delta$ given in figure (\ref{f1})  is given by $\cos\psi=g_{ij} d^i \delta^j/[\sqrt{g_{ij}d^i d^j}\sqrt{g_{ij}\delta^i \delta^j}]$ where $g_{ij}$ is the two dimensional submanifold obtained by setting $t=\mbox{constant}, \theta=\pi/2$ from GR$\Lambda$ solution of the metric (\ref{Schwarzschild}) given by
 \begin{eqnarray}\label{schdemetric}
A=B^{-1}=1-\frac{2m}{r}-\frac{\Lambda\, r^2}{3}.
 \end{eqnarray}
 Also, here $d=(dr,d\Phi)=(\beta,1)d\Phi$, with $\beta=dr/d\Phi$, is the direction of orbit of the photon whereas  $\delta=(\delta r, 0)$ is the direction of coordinate line $\Phi=\mbox{constant}$. One can obtain $\beta$ from (\ref{light-gr-soln}) as
  \begin{equation}\label{betaGR}
\beta\equiv\frac{dr}{d\Phi}=\frac{r^2}{R}\left(\frac{m}{R}\sin{2\Phi}-\cos\Phi \right).  
  \end{equation}
  Using this and GR expressions of metric (\ref{Schwarzschild}), one finds $\cos\psi= |\beta|/\sqrt{\beta^2+r^2/B}$  and from the relation $\tan\psi=\sqrt{\sec^2\psi-1}$  one finds
\begin{equation}\label{psi}
\tan\psi=\frac{r}{|\beta|\sqrt{B}}.
\end{equation}
From this expression, one can find that the one-sided bending angle is $\alpha=\psi-\Phi$ and  one can immediately calculate the bending angle for small $\Phi=\Phi_0\ll 1$. For the angle $\Phi_0$ to occur, we need from (\ref{light-gr-soln}) 
\begin{equation}\label{rrr}
\frac{1}{r}=\frac{\Phi_0}{R}+\frac{2m}{R^2},
\end{equation}
 and using this $r$ value,  from (\ref{betaGR})  one finds
 \begin{equation}
 \beta=\frac{r^2}{R}\left( \frac{2 m\, \Phi_0}{R}-1\right)\approx - \frac{r^2}{R}.
 \end{equation}
  Moreover, using the weak field GR value for $B$, whose exact form is given in equation (\ref{schdemetric}), namely  $B=1+\frac{2m}{r}+ \frac{\Lambda r^2}{3},$  and from (\ref{psi}), one finds \cite{Ishak-rindler}
\begin{equation}\label{GRdef}
\alpha_{GR\Lambda}=\frac{2m}{R}-\frac{\Lambda R\, r}{6}=\frac{2m}{R}-\frac{\Lambda R^3}{6(R\,\Phi_0+2m)},
\end{equation} 
where $r$ is given by (\ref{rrr}) and this expression reduces for $\Phi_0=0$ to the result given in \cite{Rindler-ishak} as
\begin{equation}\label{alphaGR}
\alpha_{GR\Lambda}=\alpha_E-\alpha_\Lambda, \quad\alpha_E=\frac{2m}{R},\quad \alpha_\Lambda=\frac{\Lambda R^3}{12 m}.
\end{equation}
Therefore, $\Lambda$ contributes to the deflection angle and this contribution has the opposite sign compared to the contribution due to mass of the source.  Here, we have reviewed the deflection angle for a special case where the source and observer are static. For a more general treatment of calculation of this angle when source or observer may not static, we refer to \cite{Lebedev-lake}. Having reviewed Rindler-Ishak method for GR$\Lambda$ theory, let us now apply this method to both BD$\Lambda$ and BDV theories in the following. 

\subsubsection{Deflection of light rays in BD$\Lambda$ theory}

For BD$\Lambda$ theory, from (\ref{BDBinet}),  the orbit equation becomes
  \begin{equation}\label{BDLBinetphoton}
\frac{d^2u}{d\Phi^2}+u=\frac{2m}{(2\omega+3)\phi_0 b^2}+ \frac{6 m (\omega+1)u^2}{(2\omega+3)\phi_0}- \frac{\Lambda}{3(2\omega+3) b^2 u^3},
\end{equation}
where here $b=L/E$ is a constant of motion. Most important observation of this equation is as follows. Unlike in the GR case \cite{Islam} where the corresponding equation, i.e., equation (\ref{light-gr-soln}), is independent of $\Lambda$, the last term in equation (\ref{BDLBinetphoton}) contains  $\Lambda$  explicitly. Hence $\Lambda$ clearly affects the path of photons because its solution will directly involve a cosmological constant term, even in the linear order. Note also that this term vanishes in the $\omega\rightarrow \infty$ limit and the equation reduces to the corresponding GR one given in \cite{Islam}. We will present calculation details in Appendix (\ref{appendixlight}) for clarity and here present only the results.

The solution to  equation (\ref{BDLBinetphoton}) is given in Appendix (\ref{light-soln-bdcc}). By calculating point of closest approach $r_0$ given in (\ref{pointclosestCC}) and its relation with impact parameter $b$ given in (\ref{eqclosestCC}) we see that we can use $R,r_0$ and  $b$ interchangeably in the orbit equation for linearized order and this fact enable us to simplify the solution  (\ref{light-soln-bdcc}) as
\begin{equation}\label{light-soln-bdcc-simplified}
u(\Phi)=\frac{ \sin(\Phi)}{R}+\frac{2m}{R^2 (2\omega+3)\phi_0}\left\{  1+\frac{\omega+1}{2}\left(3+\cos{2\Phi} \right) \right\} - \frac{ \Lambda R \cos^2\Phi}{6 (2\omega+3) \sin\Phi}.
\end{equation}
In comparison with GR solution (\ref{light-gr-soln}), the solution (\ref{light-soln-bdcc}) involves  $\Lambda$ explicitly. In the GR limit $\omega\rightarrow\infty$, this $\Lambda$ dependent term vanishes  and the  solution (\ref{light-soln-bdcc-simplified}) reduces to (\ref{light-gr-soln}) in GR limit.
 Hence, unlike GR$\Lambda$ case, the orbit of the photons {\bf{depends}} on $\Lambda$ and this dependence vanishes in the GR limit. 
This  implies that, in addition to the Einstein deflection angle multiplied by a multiplicative factor of $\omega$ \cite{BD}, there should be an extra contribution involving $\Lambda$ due to orbit of photons in  BD$\Lambda$ theory.  This extra deflection angle is due to the interaction of cosmological term $\Lambda$ and the scalar field and vanishes for the GR limit.
If the space time under consideration would be asymptotically flat, then one could find a deflection angle by measuring from $r\rightarrow \infty$ to find $\alpha\approx 2m/(\phi_0 R)-\Lambda R^2/[6(2\omega+3)\Phi_0]$. However, since neither the space-time is asymptotically flat nor we can ignore  effects of space-time on local measurements, we have to use an appropriate method to calculate the deflection angle. Hence, using the same method in \cite{Rindler-ishak}  we  will calculate  deflection angle and for clarity we present calculational details in Appendix (\ref{appendixlight}), see equations in appendix(\ref{betaBD}-\ref{defbdlambda}).  The result, at most  in the linear order in $m,\Phi_0$ and $\Lambda$, is 
\begin{equation}\label{defBDLambda} 
\alpha_{BD\Lambda}=\frac{2m}{\phi_0 R}-\frac{\Lambda R^2}{6(2\omega+3)\Phi_0}-\frac{\Lambda R^2}{6(2\omega+3)}\left[ (2\omega+1)\frac{r}{R}+\frac{R}{ \Phi_0^2\, r}\right], 
\end{equation}
where $r$ is given in (\ref{approx-r}). The first term in (\ref{defBDLambda}) is due to mass, the second one is the effect of $\Lambda$ on orbit and the last term is due to effect of metric on the measurement of angles.  This result reduces to special cases such as GR$\Lambda$ one (\ref{GRdef}) \cite{Ishak-rindler,Rindler-ishak,Ishak-rindler-dosset} in GR limit $\omega\rightarrow \infty$ or BD deflection angle \cite{BD,Will} for $\Lambda=0$. Also for large $\omega$ values where Cassini mission yields the gravitational deflection angle is  indistinguishable from corresponding GR$\Lambda$ expression.

 \subsubsection{Deflection of light rays in for BDV theory}
 
For this case,  from (\ref{BDBinet}), corresponding differential equation becomes
\begin{eqnarray}\label{lightdiffexp}
\frac{d^2u}{d\Phi^2}+u&=&\frac{3m u^2}{\phi_0} -\frac{2 \Lambda_0}{3(2\omega+3)b^2 u^3}
 +\frac{m\,e^{-m_s/u}}{(2\omega+3)L^2 \phi_0}\left[2E^2-3 L^2 u^2+O(m_s,m_s^2)  \right].
\end{eqnarray}
Due to the exponential term, this equation is complicated. However, this equation can be analyzed for very massive or  light scalar cases as follows:
\begin{itemize}
\item{}For a  very massive scalar, $m_s\gg 1$,  the exponential term can be ignored and together with $\phi_0=1$ for this case, the equation (\ref{lightdiffexp}) reduces to
\begin{eqnarray}
\frac{d^2u}{d\Phi^2}+u&=&3m u^2 -\frac{2 \Lambda_0}{3(2\omega+3) b^2 u^3},
\end{eqnarray}
 which is exactly the same with GR case except for the last term in the equation.  Hence the linearised solution would be a mixture of GR and BD$\Lambda$ solutions given by
\begin{equation} 
u(\Phi)=u_{GR}(\Phi)-\frac{ \Lambda_0 R^3 \cos^2{\Phi} }{3 \, b^2\, (2\omega+3)\sin{\Phi}},
\end{equation}
where  $u_{GR}$ is solution of GR case given in (\ref{light-gr-soln}).
Thus, the minimum of potential, $V_0=2\Lambda_0$, enters in the orbit equation and will affect the light deflection. Repeating similar calculations, one finds that the bending angle in the linearised order become
\begin{equation}\label{defbdpotmass}
\alpha_{BDmassive}=\alpha_E-\frac{\Lambda_0 R^2}{3(2\omega+3)\Phi_0}-\frac{\Lambda_0 R^2}{3(2\omega+3)}\left[ \frac{(2\omega-1)r}{2R}+\frac{R}{\Phi^2_0\, r}\right]. 
\end{equation} 
 Here $r$ is given by  (\ref{approx-r}) with $\Lambda$ to be replaced by $2\Lambda_0$ 
 together with  $\phi_0=1$. Therefore, for a very massive scalar  field, as it is well known, light deflection due to mass is exactly the same with GR, and the effect of minimum of the potential acting as a cosmological constant has  a slightly different $\omega$ dependence compared with the result of  BD$\Lambda$ theory given in (\ref{defBDLambda}).  
 
\item{} For a very light scalar, $m_s\ll 1$, we can expand the exponential term in equation (\ref{lightdiffexp}) and find the following equation:
\begin{eqnarray}
\frac{d^2u}{d\Phi^2}+u&=&\frac{2m}{(2\omega+3)\phi_0 b^2}+\frac{6m (\omega+1) u^2}{(2\omega+3)\phi_0} -\frac{2\Lambda_0}{3(2\omega+3)b^2 u^3}+O(m\times m_s).
\end{eqnarray}
Here we see that the resulting differential equation resembles the same form with  the  equation (\ref{BDLBinetphoton}) of the BD$\Lambda$ case except $\Lambda$ is replaced by $2\Lambda_0$. Therefore its linearised solution (\ref{light-soln-bdcc}) will have same form except $\Lambda$ to be replaced by $2\Lambda_0$. The effect of minimum of the potential acting as a cosmological constant defined by $V_0=2\Lambda_0\phi_0$ on the total deflection angle can be seen from the total deflection angle expression given by
\begin{eqnarray}\label{deflightbdpotlight}
\alpha_{BDV-lightsc.}&=&\frac{2m}{\phi_0\, R}-\frac{\Lambda_0 R^2}{3(2\omega+3)\Phi_0}- \frac{\Lambda_0 R^2}{3(2\omega+3)}\left[ \frac{(2\omega-1) r}{2R} +\frac{R}{\Phi_0^2 \,r}\right].
\end{eqnarray}
Here $r$ is given by (\ref{approx-r}) with $\Lambda$ to be replaced by $2\Lambda_0$.  
In comparison with the result of BD$\Lambda$ theory given in (\ref{defBDLambda}), there are some slight differences for  corresponding results of  both light scalar (\ref{defbdpotmass}) and massive scalar (\ref{deflightbdpotlight}) cases. These differences originate from small differences of the metric functions $A$ and $B$ for BD$\Lambda$ and BDV theories. Due to observational results, the very light scalar mass case cannot deviate from corresponding GR$\Lambda$ expression since $\omega>40.000$ limit is also valid for this case. But for BDV theory with very massive scalar field,  there is no such limit on $\omega$ and the deflection angle due to minimum of potential can be very different from corresponding GR$\Lambda$ expression. However, there is no observational data measuring the deflection angle due to the cosmological constant, yet. Hence, there is a possibility  that can limit the parameters of these theories or eliminate them  if such an observation is made in the future.      
 
 \end{itemize}
 
 \section{Gravitational Redshift}
 
 The spacetime described by (\ref{Schwarzschild}) and (\ref{Schmetric}) is stationary. Hence it admits a timelike Killing vector. In these coordinates,  the ratio of the measured frequency $\nu$ of a light passing through different positions is given by 
 \begin{equation}\label{redshift}
\frac{\nu_0}{\nu}=\sqrt{\frac{A(r)}{A(r_0)}} 
 \end{equation}  
Reading metric function $A(r)$ from (\ref{Schmetric}) and considering the fact we are working in the linear level,  the equation (\ref{redshift}) becomes
\begin{equation}
\frac{\nu_0}{\nu}=1+\frac{m}{\phi_0 r_0}-\frac{m}{\phi_0 r}-\frac{\Lambda}{6}\left(r^2-r_0^2 \right)+\frac{\varphi(r_0)-\varphi(r)}{2\phi_0}.
\end{equation}
In the GR limit this expression reduces to the one given in \cite{Kagramanova}. The effects of the scalar field to the gravitational redshift is given by the last term. Let us evaluate these for BD$\Lambda$ and BDV theories, separately.
\subsection{Gravitational redshift for BD$\Lambda$ case}
For this case, reading $\varphi$ from (\ref{varphiBDL}),  and $\phi_0$ from (\ref{phi01}) we find
\begin{equation}
\frac{\nu_0}{\nu}=1+\frac{m}{r_0}-\frac{m}{r}-\frac{2\omega+2}{2\omega+3}\frac{\Lambda}{6}\left(r^2-r_0^2 \right),
\end{equation}
hence the gravitational redshift due to mass is the same as in GR whereas there is a correction term involving BD parameter $\omega$ for the gravitational redshift due to the cosmological constant term. Since the result of  Cassini mission requires large $\omega$, this factor approaches to one and the expression becomes identical to GR one for $BD\Lambda$ theory.  

\subsection{Gravitational redshift for BDV case}
For this case, by considering  (\ref{varphiBDV}) we have
\begin{equation}
\frac{\nu_0}{\nu}=1+\frac{m}{\phi_0 r_0}\left(1+\frac{e^{-m_s r_0}}{2\omega+3} \right)-\frac{m}{\phi_0 r}\left(1+\frac{e^{-m_s r}}{2\omega+3}  \right)-\frac{2\omega+1}{2\omega+3}\frac{V_0}{12 \phi_0}\left(r^2-r_0^2 \right).
\end{equation}
Hence  the gravitational redshift due to mass term  is modified  since each term is multiplied by a position dependent effective gravitational term. The term for the minimum of the potential has a similar contribution but the multiplicative factor involving BD parameter is slightly different than BD$\Lambda$ case. We can expand the  terms involving  mass terms for a very light or very heavy scalar field mass cases as follows.
\begin{itemize}
\item{} For a very heavy scalar, $m_s\rightarrow \infty$, since $e^{-m_s r}\rightarrow 0$ and $\phi_0=1$  we find that
\begin{equation}
\frac{\nu_0}{\nu}=1+\frac{m}{r_0}-\frac{m}{r}-\frac{2\omega+1}{2\omega+3}\frac{V_0}{12}\left(r^2-r_0^2 \right).
\end{equation}
Similar to the previous case, the gravitational redshift due to mass is the same as in GR and the term due to minimum of the potential gets a multiplicative factor, same as in advance of perihelion for this theory. For $BDV$ theory with heavy scalar field mass, there is no lower limit for the value of $\omega$. Hence if we regard $V_0$ as a cosmological constant using  the equation $V_0=2\Lambda_0$, depending on the value of $\omega$, the redshift term due to $V_0$ can take any value in the interval $(-\infty,\infty)$  which can be seen from the behavior of multiplicative factor given in figure (\ref{fig1}). Using the argument given in \cite{Kagramanova} which considers the result of Gravity Probe-A experiment \cite{gpa} one can put a bound $
\left|(2\omega+1)\Lambda_0/(2\omega+3) \right|\le 10^{-28} m^{-2}$ but this bound is much larger than the current value of cosmological constant. If  future experiments  will reach enough sensitivity, then one can use this phenomena to restrict the parameter space $(\omega,\Lambda_0)$ of this theory.

\item{} For a very light scalar, $m_s\rightarrow 0$, $e^{-m_s r}\sim 1-m_s r$, ignoring multiplication of $m$ with $m_s$ and using  the value of $\phi_0$ given in equation (\ref{phi01}) for this case, we find
\begin{equation}
\frac{\nu_0}{\nu}=1+\frac{m}{r_0}-\frac{m}{r}-\frac{2\omega+1}{2\omega+3}\frac{V_0}{12\phi_0}\left(r^2-r_0^2 \right).
\end{equation}
Here again  the gravitational redshift due to mass is the same with GR case \cite{Kagramanova}, and the gravitational redshift due to the minimum of the potential contains a numerical factor involving $\omega$ similar to heavy scalar field mass case. However, unlike from heavy case, for a very light scalar, this factor approaches to one since Cassini mission requires  $\omega\ge 40.000$ for light BDV theory. Hence, there is no deviation from GR$\Lambda$ results for this case.
\end{itemize} 

\section{Galaxy dynamics}
In GR it is well known that the effects of the cosmological constant or dark energy on the solar system scales or galactic scales are too weak to be observable. However, when the scales comparable or bigger than 1 Mpc, its effects cannot be ignorable anymore. Here, with the help of using the results we have obtained in the previous sections, we will discuss the effects of cosmological constant or minimum of the potential of BD$\Lambda$ and BDV theories on the local dynamics of the universe and whether the results agree with GR.

\subsection{Galaxy rotation curves}
To discuss these effects in the galactic scale, we can consider galaxy rotation curves. It was observed \cite{Rubin1,Rubin2} that the rotation curves of gas  at the outer regions of galaxies show a nearly constant  velocity up to several galactic luminous radii. To apply our results to this phenomena, now,  first let us calculate the rotational velocity of stars around the center of a  static, spherically symmetric galaxy. We can express radial geodesics equation  on equatorial plane  (\ref{radialeom}) for timelike particles  as 
\begin{equation}
\dot{r}^2+U(r)=0,
\end{equation}
 where 
\begin{equation}
 U(r)=\frac{1}{B}\left(1-\frac{E^2}{A}+\frac{L^2}{r^2} \right).
 \end{equation}
The conditions for the existence of stable circular orbits  are:
\begin{eqnarray}\label{stablecircular}
\dot{r}=0\quad (U(r)=0),\quad U'(r)=0,\quad U''(r)>0.
\end{eqnarray} 
Here $'$ denotes derivative with respect to $r$. From the first two conditions with a little algebra one finds
\begin{eqnarray}
E^2=\frac{2A^2}{2A-rA'},\quad
L^2=\frac{r^3 A'}{2A-rA'}. \label{EA}
\end{eqnarray}
Moreover, the second derivative of the potential becomes
\begin{equation}
U''=\frac{2}{r B}\left[\frac{r A''+A'(3-2 r A')}{2A-rA'}\right].
\end{equation}
The above conditions were already obtained in the previous works, for example in \cite{Matos}.  A numerical investigation showed that the last condition in (\ref{stablecircular}) is satisfied in the relevant values of $r$. From the proper time expression $d\tau^2=-ds^2$, considering the definition of four velocity $U^{\mu}=dx^\mu/d\tau=(\dot t,\dot r,\dot \theta=0,\dot \Phi)$ we find that
\begin{equation}
1=A (U^0)^2\left(1-v^2\right),
\end{equation}
where $U^0$ is the time component of the four velocity of the particle  and $v$ is the spatial velocity defined as
\begin{equation}
v^2=\frac{1}{A}\left[B \left(\frac{dr}{dt}\right)^2+r^2 \left(\frac{d\Phi}{dt} \right)^2 \right]=(v^r)^2+(v^\phi)^2,
\end{equation}
where $v^r$ and $v^\phi$ are the  components of the spatial velocity $v$ which is observed in an orthonormal coordinate system. Its $\Phi$ component is given by
\begin{equation}
v^\Phi=\frac{r}{\sqrt{A}}\Omega, \quad \Omega=\frac{d\Phi}{dt},
\end{equation}
From the first integrals of the geodesics equation and  Eq.(\ref{EA}) we can calculate $\Omega$ as
\begin{equation}
\Omega=\frac{d\Phi}{dt}=\frac{\dot{\Phi}}{\dot{t}}=\frac{A}{r^2}\frac{L}{E}=\sqrt{\frac{A'}{2r}}.
\end{equation}
Using this value, we find the tangential velocity of a particle in a stable circular motion as follows
\begin{equation}\label{tangentialvelocity}
(v^\Phi)^2=\frac{r A'}{2A}.
%=\left(\frac{\frac{m}{\phi_0 r} - \frac{\Lambda r^2}{3}-\frac{r\,\varphi'}%{2\phi_0}}{1-\frac{2m}{\phi_0 r}-\frac{\Lambda r^2}{3}-\frac{\varphi}%{\phi_0}}\right)^{1/2}.
\end{equation}
In the linearized approximation, we find that
\begin{equation}
(v^\Phi)^2=\frac{m}{\phi_0 r} - \frac{\Lambda r^2}{3}-\frac{r\,\varphi'}{2\phi_0}.
\end{equation}
Thus, the effects of the BD scalar field reveals itself in the last term as well as in the constant $\phi_0$ for this phenomena. Let us discuss this term for the theories we consider separately.
\begin{itemize}
\item{} For BD$\Lambda$ theory we have
\begin{equation}\label{vphibdl}
(v^\Phi)^2_{BD\Lambda}=\frac{m}{r} - \frac{2\omega+2}{2\omega+3}\frac{\Lambda r^2}{3}.
\end{equation} 
In order that these expression can have somewhat constant behaviour, the sign of the last term after the minus sign must be negative, which is possible for the interval $-3/2< \omega < -1$ for positive $\Lambda$, as can be seen in figure (\ref{fig1}). This is however ruled of by the result of the Cassini mission with the requirement that $\omega>40.000$. Hence, the cosmological constant term cannot explain the flat rotation curves of galaxies for $BD\Lambda$ theory. 

\item{}For BDV theory we find that 
\begin{equation}\label{vphibdv}
(v^\Phi)^2_{BDV}=\frac{m}{\phi_0 r}\left[1+\frac{e^{-m_s r}}{(2\omega+3)}\left(m_s r+1 \right) \right] - \frac{2\omega+1}{2\omega+3}\frac{V_0\, r^2}{6\phi_0}.
\end{equation} 

\end{itemize}

We can again look for special cases for this expression. For a heavy scalar, exponential term vanishes and  the first term in tangential velocity becomes similar to BD$\Lambda$ case since $\phi_0=1$.    For a very light scalar the term involving mass  becomes again the same as in (\ref{vphibdl}) and for the term containing $V_0$ we have to take $\phi_0$ as in (\ref{phi01}). In all these cases we see that the numerical factor involving $\omega$ does not change the order of magnitude of the term related to $\Lambda$ or $V_0$,  for positive $ \omega$.
Hence, these terms cannot explain flat rotation curves of stars in a galaxy for positive $\omega$. For negative values of $\omega$, the factors containing $\omega$ may be negative and there may be regions where nearly flat rotation curves possible in principle. However, similar to  $BD\Lambda$ theory, $BDV$ theory with light scalar mass, solar system tests require large positive values of $\omega$ and this posibility is ruled out. For a heavy scalar field mass, however, there is no restriction on $\omega$ by solar system tests and for $-3/2<\omega <-1/2$, the coefficient of the last term of (\ref{vphibdl}) after minus sign becomes negative  for positive $V_0$, making this term an increasing function of $r$. Hence, for this range, the minimum of potential can contribute to the flat  rotational curves of galaxies. For the values of $\omega$ outside this range, however, the minimum of the potential cannot contribute to  flat rotation curves. In GR the flat rotation curves is explained by the existence of dark matter, usually modeled as a  dust or perfect fluid surrounding the galactic core which interacts with other particles  only via gravity. This behaviour can also  be explained by the existence of exotic sources such as a global monopole  behaving as a galactic dark matter \cite{Nucamendi,Hoonlee}. It might be interesing to consider a dust or perfect fluid source for $BD\Lambda$ and $BDV$ theories as a candidate of dark matter. We are  currently working on this problem and we will  present our results elsewhere.  

For intermediate values of $m_s$ where these approximations are not valid, the Yukawa type term  in (\ref{vphibdv}) may  also explain the flat rotation curves. This is because Sanders showed  in \cite{Anders} that  a Yukawa type  phenomenical gravitational potential can explain the behavior of galaxy rotation curves. In that work  the following expression for rotational velocity is obtained 
\begin{equation}\label{vphibdvAnders}
(v^\Phi)^2=\frac{G_\infty m}{ r}\left[1+\alpha e^{-\frac{r}{r_0}}\left( \frac{r}{r_0}+1 \right) \right].
\end{equation} 
In this expression $G_\infty$ is the gravitational constant measured at infinity, $r_0$ is a length scale of this potential and $\alpha$ is a coupling constant of this Yukawa type term. Sanders showed that in the presence of a Yukawa type gravitational potential, for $-0.95<\alpha<-0.92$ there is a region where the general properties of extended galactic rotation curves are reproduced.    Comparing our expression (\ref{vphibdv}) with (\ref{vphibdvAnders}), we see that they are similar if we identify $\alpha=(2\omega+3)^{-1}$, $r_0=1/m_s$, $G_\infty=1/\phi_0$. Then, the above limit on $\alpha$ is equivalent to $-2.04<\omega<-2.02$. Hence, a generic BDV theory can explain the observed galaxy rotation curves without needing a dark matter if BD parameter $\omega$ is in this interval. This result is also discussed in \cite{Stabile} for a generic  $f(R,\phi)$ gravity including BDV theory as  a special case. The problem here is that the ranges of $\omega$ where the observed galaxy rotation curves were reproduced are very restricted negative and unfavourable values  of it. 
 The contribution of minimum of the potential to rotational velocity  is in the reducing sense since the multiplicative factor is positive for this value of $\omega$. In summary for a very restricted and negative value of $\omega$, rotational curves of galaxies can be explained by the mass of the scalar field of BDV theory leading to a Yukawa type term.  
 
\subsection{Inter-Galactic dynamics}
Now let us turn our attention to inter-galactic scales. By using the radial geodesics equation $\ddot{r}+\Gamma^r_{\mu\nu}\dot{x}^\mu\dot{x}^{\nu}=0$, and the first integrals of motion given in (\ref{firstintegrals}), we find an equation describing the radial acelerations of galaxies towards each other as
\begin{equation}\label{radac}
\frac{d^2r}{dt^2}=-\frac{A'}{2B}=-\frac{m}{\phi_0 r^2}+\frac{\Lambda\, r}{3}+\frac{\varphi'}{2\phi_0},
\end{equation}    
where $r$ describes radial separation between two galaxies and $m$ is the total mass of the galaxies. Here inner structures and relative rotations of galaxies are ignored, merely by treating them as two point particles. Now we evaluate this equation for the theories we are considering in this paper. 

\begin{itemize}

\item{}  For BD$\Lambda$  theory, the aceleration equation (\ref{radac}) has the form
\begin{equation}\label{radacccc}
\frac{d^2r}{dt^2}=-\frac{m}{ r^2}+\frac{2\omega+2}{2\omega+3}\frac{\Lambda r}{3},
\end{equation}
and the only difference with respect to corresponding GR$\Lambda$ expression \cite{Lahav}  is the factor involving $\omega$ in front of the cosmological constant. In order to better understand the effects  of the mass and cosmological constant on the  dynamics of the galaxies let us calculate the ratio of both terms in   
(\ref{radacccc}) and denote by q, which is given by
\begin{equation}
q_{BD\Lambda}=\frac{2\omega+2}{2\omega+3}\frac{\Lambda r^3}{3m}=\frac{2\omega+2}{2\omega+3}q_{GR\Lambda}.
\end{equation}
where at the last step the corresponding $GR\Lambda$ expression of $q$,  discussed in detail in \cite{Axenides} is identified. Hence, it is clear that the difference between corresponding equation of GR$\Lambda$ theory is the  factor involving $\omega$.  Let us now discuss the behavior of this factor.  This factor has the range $(2/3,1)$ for $\omega>0$,  hence it does not change the   order of $q$ in this range of $\omega$. The behaviour of this factor  is more complicated for negative values of $\omega$, which can be seen at the graph (\ref{fig1}). This factor even vanishes for $\omega=-1$ where  the cosmological constant has no effect on galaxy dynamics. It even takes  negative values  in the range $-3/2\le \omega \le -1, $ where in this range the effect of positive cosmological constant is attractive rather than repulsive. However, the value of $\omega$ is fixed by solar system tests as $\omega \ge 40.000$ and the dramatic changes of the behaviour of $\Lambda$ for  negative values of it is ruled out.  Therefore, we can have similar conclusions given in \cite{Axenides},  namely if we take the value of  $\Lambda$  as its the recent observed value,  then cosmological constant does not affect interplanetary and galactic scales and its effects  becomes significant at the cluster scales for BD$\Lambda$ theory. This is because for solar system $q_{GR\Lambda}\sim 10^{-20}$, for galactic scale $q_{GR\Lambda}\sim  10^{-4}$ but for cluster scale $q_{GR\Lambda}\sim O(1)$ \cite{Axenides}.  However, for the theories where this factor is not fixed by solar system tests, these extreme behaviors can be still possible.

\item{} For BDV  theory the acceleration equation (\ref{radac}) becomes
\begin{equation}\label{radacccc}
\frac{d^2r}{dt^2}=-\frac{m}{\phi_0 r^2}\left[ 1+\frac{(1+m_s r)e^{-m_s r}}{2\omega+3}\right]+\frac{2\omega+1}{2\omega+3}\frac{V_0 r}{6 \phi_0},
\end{equation}
including a Yukawa like term in the expression. For the case where $m_s\gg 1$ we have $e^{-m_s}\rightarrow 0$ and $\phi_0=1$, we have

\begin{equation}\label{radaccccmassive}
\frac{d^2r}{dt^2}=-\frac{m}{ r^2}+\frac{2\omega+1}{2\omega+3}\frac{V_0 r}{6 }.
\end{equation}
Hence for this case the mass term is the same with the GR case and the term related to the minimum of the potential has the same factor involving $\omega$ as in other cases of this theory. 

For a very light scalar, $m_s\ll 1$,we can ignore terms involving $m\times m_s$ and we find 
\begin{equation}\label{radacccclight}
\frac{d^2r}{dt^2}=-\frac{m}{ r^2}+\frac{2\omega+1}{2\omega+3}\frac{V_0 r}{6 \phi_0}.
\end{equation}
Here again the term containing $V_0$ has a  numerical factor involving $\omega$,  different from both BD$\Lambda$  and massive $BDV$ cases. For both theories the  q factor becomes
\begin{equation}\label{qbdv}
q_{BDV}=\frac{2\omega+1}{2\omega+3}\frac{V_0 r^3}{6\phi_0 m}.
\end{equation}
For a light scalar, the value of $\omega$ is fixed by solar system tests as $\omega>40.000$ and the numerical factor involving $\omega$ of (\ref{qbdv})  has no effect. Hence, similar to BD$\Lambda$ theory, the effects of the cosmological constant becomes relevant at the cluster scales for BDV theory with very light scalar field mass. For very heavy scalar field mass case, however, solar system tests do not fix the value  of $\omega$ and the numerical factor may become important for small or negative values of it.  This fact may have two consequences for BD$V$ theory with heavy scalar field mass: 
\item{1)}Since $\omega$ is not fixed, significant deviations from the results of GR$\Lambda$ theory can be possible to observe in principle for negative values of $\omega$, since the behaviour of the factor for negative values of $\omega$ may be quite large as seen in the figure (\ref{fig1}).  
\item{2)}  Phenomena at ranges larger then solar system scale  may  help to limit BD parameter $\omega$ for this theory compared to GR$\Lambda$ theory if independent measurements determine the value of $V_0$ and $m$ in (\ref{qbdv}).  
 \end{itemize}
 
\section{Conclusions}

In this paper, we have discussed weak field equations of Brans-Dicke scalar-tensor theory extended by the presence of either a cosmological constant term coupled linearly to the scalar field or a generic potential in the Jordan frame. The linearized  field equations of both cases are obtained in the gauge choice which makes the scalar field terms decouple from the metric field equations. The most important differences of both theories is that the former leads to a massless scalar field with a source whereas the latter has a massive scalar field where mass term is proportional to second derivative of the potential in the Taylor expansion as usual. To our knowledge, the linearized expansion of the former case is not present in the literature. 

In the second part of the paper, we have considered the  weak field solutions  for a massive point particle  for both theories in the linear approximation. The solutions have been first obtained in the gauge  employed and then transformed to some physically relevant coordinates such as isotropic and Schwarzschild type coordinates. As a physical application, particle motion of test particles has been investigated  with the focus on the solar system effects such as advance of perihelion,  deflection of light rays and gravitational redshift. The effect of mass of the source, cosmological term or minimum of the potential on these phenomena were derived in the linear order. The effect of the mass of the scalar field is also determined for $BDV$ theory,  which contains Yukawa like terms,  but analytic solutions were derived only for very light or very massive scalar field. The effects of the terms responsible for asymptotical nonflatness, namely $\Lambda$ or $V_0$ are similar to cosmological constant in GR$\Lambda$ theory except some factors involving BD parameter $\omega$, which are different for both theories. This might imply a new observational window in the future, for example to limit $\omega$ for BD$\Lambda$ or BDV theories. However, the Casini mission limits BD parameter to $\omega >40.000$ \cite{cassini} for original BD theory, and this limit is also valid for BD$\Lambda$ theory and BDV theory with very light scalar. Hence, we conclude that the effects of the cosmological constant or the minimum of the potential are  indistinguishable for these theories. For BDV theory with a very heavy mass,  however, since the scalar field has a very short range and freezes out outside this range, the effect of mass of the source to this phenomena becomes identical to corresponding GR one. Hence solar system test are satisfied irrespective of the value of $\omega$. Therefore, the correction to these phenomena due to the minimum of potential has a factor involving $\omega$, whose value can take much larger and smaller values then $O(1)$ as given in figure (\ref{fig1}). Hence, for BDV theory with a very heavy mass, the effect of minimum of potential may be different from corresponding GR one even if one uses  the same observed value of cosmological constant for the minimum of the potential. This fact may even lead to put some bounds on $\omega$ for very massive $BDV$ theory if  these  phenomena  will be measured with enough sensitivity in the future. 

The latter part of this work is devoted to galactic and intergalactic dynamics of these theories. For the galaxy scale we have calculated rotational velocity of stars in a galaxy and see that the nearly flat region of the galaxy rotation curve cannot be explained by the cosmological constant of BD$\Lambda$ theory as well as the minimum of potential for BDV theory with light scalar field mass. Moreover, the effects of mass and the cosmological constant or minimum of potential becomes indistinguishable from corresponding GR ones since the factors involving $\omega$ becomes equal to one for the observed limit of $\omega$.  For a very heavy scalar field mass, however, since there is no limit on $\omega$ due to solar system tests, there is a range of $\omega$ where the correction factor becomes negative. Hence the contribution of the minimum of potential becomes an increasing function of $r$,  which may contribute to the flat rotation curves for $-3/2<\omega<-1/2$. Outside this range the minimum of the potential cannot contribute to such behavior for BDV theory with a very heavy scalar field mass. For generic values of the mass of the scalar field,  the  flat rotation curves can also be explained by the effect of the mass of the scalar field for a very limited negative range of $\omega$. This is because the mass of the scalar  field  introduces a Yukawa like term in rotation velocity expression and this term can produce such a behaviour if $-2.04< \omega< -2.02$.   For the intergalactic scale, we have generalized the GR$\Lambda$ expression corresponding to the acceleration of two galaxies towards each other where we have treated galaxies as point particles.   We have obtained a  factor $q$ which can determine the scale where the contribution of the cosmological constant starts to become relevant when this factor becomes of the order of unity. Similar to other phenomena we have discussed, this factor becomes indistinguishable for BD$\Lambda$ or BDV theory with a very light scalar mass  from corresponding GR$\Lambda$ case, due to the large value the solar system tests sets on  the BD parameter $\omega$. For BDV theory with heavy scalar mass,  the scale where the  factor $q$ becomes at the order of unity  can be very different than corresponding GR$\Lambda$ theory even if we use the minimum of potential equal to the observed value of the cosmological constant. Hence, these phenomena may lead to test the BDV theory with a very heavy scalar field mass or to limit the range the parameter $\omega$  compatible with observations if in the future there will be  observations with enough sensitivity to  determine the other parameters of the theory.

\section*{Acknowledgements}
H.O. is partially supported by TUBITAK 2211/E Programme.
\appendix

\section{Deflection of light rays in BD$\Lambda$ theory\label{appendixlight}}

Here let us find a solution to the orbit equation (\ref{BDLBinetphoton}) in the linearized level.  In order to find the effect of the point mass and cosmological constant on a light ray coming from very far region of spacetime, we consider a perturbative approach, and consider the following ansatz
\begin{equation}\label{photonans}
u(\Phi)=u_0(\Phi)+m u_1(\Phi)+\Lambda u_2(\Phi).
\end{equation}  
Replacing the (\ref{photonans}) into  equation (\ref{BDLBinetphoton}) we obtain following set of equations in the zeroth order, orders linear on $m$ and  $\Lambda$ as
\begin{eqnarray}
&&u_0''+u_0=0,\\
&&u_1''+u_1=\frac{2}{b^2 (2\omega+3)\phi_0}+\frac{6 (\omega+1) u_0^2}{(2\omega+3)\phi_0},\\
&&u_2''+u_2=-\frac{1}{3 b^2 (2\omega+3)u_0^2}.
\end{eqnarray}
The solution of the first equation yields a photon following a straight line with
\begin{equation}
u_0(\Phi)=\frac{ \sin(\Phi)}{R}.
\end{equation}
Replacing this into remaining equations, one obtains the solution
\begin{equation}\label{light-soln-bdcc}
u(\Phi)=\frac{1}{r(\Phi)}=\frac{ \sin(\Phi)}{R}+m\left\{\frac{2}{b^2(2\omega+3)\phi_0}+\frac{(\omega+1)[3+\cos{(2\Phi)}]}{(2\omega+3)\phi_0 R^2} \right\} - \frac{ \Lambda R^3 \cos^2\Phi}{6b^2 (2\omega+3) \sin\Phi}.
\end{equation}
In comparison with GR case, the solution (\ref{light-soln-bdcc}) involves $b$ and $\Lambda$ and these parts vanish in the GR limit $\omega\rightarrow\infty$.
This solution implies that, in addition to the Einstein deflection angle multiplied by a multiplicative factor of $\omega$, an extra contribution comes from the cosmological term.  This extra deflection angle is due to the interaction of cosmological term $\Lambda$ and the scalar field and vanishes for the GR limit. Hence, unlike GR case, the orbit of the photons {\bf{depend}} on $\Lambda$. 
Note that the relation between integration constant $R$ and closest approach distance $r_0$ can be found by setting $\Phi=\pi/2$ in (\ref{light-soln-bdcc}). This yields
\begin{equation}\label{pointclosestCC}
\frac{1}{r_0}=\frac{1}{R}+\frac{2m}{(2\omega+3)\phi_0}\left[\frac{1}{b^2}+\frac{(\omega+1)}{ R^2} \right].
\end{equation}
 The equation satisfied by closest approach distance $dr/d\Phi=0$ is given by
 \begin{equation}\label{eqclosestCC}
 \frac{1}{b^2}+\frac{(2\omega+2)\Lambda}{3(2\omega+3)}=\frac{1}{r_0^2}-\frac{2m \,(2\omega+4)}{(2\omega+3)\phi_0\, r_0^3}.
 \end{equation}
These relations show that in the linearized order in the orbit equation (\ref{light-soln-bdcc}) we can take $R=r_0=b$ interchangeably. Thus the solution can be expressed in terms of only one of these constants, such as $R$.   Using this fact, the solution (\ref{light-soln-bdcc}) simplifies to (\ref{light-soln-bdcc-simplified}). 

In order to calculate the deflection angle, we use the method developed in \cite{Rindler-ishak} by Rindler and Ishak.  The deflection angle can be calculated from (\ref{psi}) 
 where here 
\begin{equation} \label{betaBD}
 \beta=\frac{dr}{d\Phi}=\frac{m r^2}{R^2}\frac{2(\omega+1)}{(2\omega+3)\phi_0}\sin 2\Phi-\frac{r^2}{R}\cos\Phi-\frac{\Lambda R r^2}{6 (2\omega+3)}\cos \Phi\left( 1+\frac{1}{\sin^2 \Phi}\right).
 \end{equation}
Here, unlike \cite{Rindler-ishak}, there is singularity in the solutions for $\Phi=0$, so we measure the deflection angle at $\Phi=\Phi_0\ll 1 $ where deflection for mass is already achieved. We will use small angle approximations $\sin\Phi_0\approx \Phi_0$,  $\cos\Phi_0\approx 1.$   
 Then, from (\ref{light-soln-bdcc}) we have
 \begin{equation}\label{approx-r}
u=\frac{1}{r}\approx \frac{\Phi_0}{R}+\frac{2m}{\phi_0 R^2} -\frac{\Lambda R }{6 (2\omega+3) \Phi_0}.
 \end{equation}
 The value of $\beta$  in (\ref{betaBD}) becomes  
\begin{eqnarray}\label{beta1}
|\beta|=\frac{r^2}{R}\left[1- \frac{2m(2\omega+2)}{(2\omega+3)\phi_0 R}\Phi_0+\frac{\Lambda R^2}{6(2\omega+3)}\left(1+\frac{1}{\Phi_0^2} \right)\right].
\end{eqnarray} 
Note that the angle $\Phi_0$ should be at the same order of magnitude as the other parameters in (\ref{light-soln-bdcc}), namely we can choose $\Phi_0\approx O(m/R)$.

Using these results and the metric function $B$  evaluated at the  $r$ value (\ref{approx-r}), and $\beta$ given in (\ref{beta1}), the expression (\ref{psi}) yields the following result, at most the linear order of the parameters $m,\Lambda$ and $\Phi_0$, as  
\begin{eqnarray} \label{defbdlambda}
 \psi&=&\frac{r}{|\beta|\sqrt{B}}=\frac{R}{r}\left[1- \frac{2(2\omega+2)m \Phi_0}{(2\omega+3)\phi_0 R}+\frac{\Lambda R^2}{6(2\omega+3)}\left(1+\frac{1}{\Phi_0^2} \right)\right]^{-1}
 \left[1-\frac{(2\omega+2)\,m}{(2\omega+3)\phi_0 r}-\frac{(2\omega+1)\Lambda r^2}{6(2\omega+3)} \right] \\
 &\approx & \Phi_0+\frac{2m}{\phi_0 R}-\frac{\Lambda R^2}{6(2\omega+3)\Phi_0}-
\frac{ (2\omega+1)\Lambda R^3}{ 6(2\omega+3)\left[\Phi_0 R+\frac{2m}{\phi_0}-\frac{\Lambda R^3}{6(2\omega+3)\Phi_0} \right]}-\frac{\Lambda R^3}{6(2\omega+3)\Phi_0^2}\left[\frac{\Phi_0 R+\frac{2m}{\phi_0}-\frac{\Lambda R^3}{6(2\omega+3) \Phi_0}}{R^2} \right].\nonumber
\end{eqnarray} 
In deriving this we have supposed that $\Lambda$ is much smaller than the other parameters $m$ and $\Phi_0$. This expression yields our result given in equation (\ref{defBDLambda}) for half deflection angle for BD$\Lambda$ theory defined as $\alpha=\psi-\Phi_0$.

\end{document}